\documentclass[showpacs,preprintnumbers,amsmath,amssymb,nofootinbib]{revtex4}

\usepackage{graphicx,bm}
\usepackage{dcolumn}    

\newcommand{\hatrho}{{\hat{\rho}_0}}
\newcommand{\rhomax}{2}

\begin{document}

\title{ 
Six-dimensional localized black holes: numerical solutions 
}
\author{Hideaki Kudoh } 
\affiliation{   Department of Physics, Kyoto University, Kyoto 606-8502, Japan  } 

\begin{abstract}
To test the strong-gravity regime in Randall-Sundrum braneworlds, we consider black holes bound to a brane. 
In a previous paper, we studied numerical solutions of localized black holes whose horizon radii are smaller than the AdS curvature radius.
In this paper, we improve the numerical method and discuss properties of the six dimensional (6D) localized black holes whose horizon radii are larger than the AdS curvature radius. 
At a horizon temperature $\mathcal{T} \approx 1/2\pi \ell$, the thermodynamics of the localized black hole undergo a transition with its character changing from a 6D Schwarzschild black hole type to a 6D black string type. 
The specific heat of the localized black holes is negative, and the entropy is greater than or nearly equal to that of the 6D black strings with the same thermodynamic mass.
The large localized black holes show flattened horizon geometries, and the intrinsic curvature of the horizon four-geometry becomes negative near the brane. 
Our results indicate that the recovery mechanism of lower-dimensional Einstein gravity on the brane works even in the presence of the black holes. 
\end{abstract}


\preprint{KUNS-1897}
\preprint{hep-th/0401229}


\maketitle

%

\section{Introduction}

Much progress has been made in understanding the dynamics of gravity, brane and localized matter fields in the context of braneworlds. 
The braneworld models proposed by Randall and Sundrum (RS) \cite{Randall:1999ee,Randall:1999vf} have an extra dimension that is compactified via anti-de Sitter (AdS) space and branes, assuming that we live on a brane (or a domain wall) inside the AdS space. This compactification is called warped compactification. The RS models do not require a large extra dimension due to the warped compactification, contrary to other interesting braneworld models with large extra dimensions \cite{Arkani-Hamed:1998rs,Antoniadis:1998ig}.

In the warped compactification, analyses of linear and second-order perturbations show that gravity induced on the brane by localized matter fields is consistent with usual four-dimensional Einstein gravity~\cite{Shiromizu:1999wj,Sasaki:1999mi,Garriga:1999yh,Tanaka:2000er,
Mukohyama:2001jv,Mukohyama:2001ks,Giannakis:2000zx,Kudoh:2001wb,
Kudoh:2001kz,Kudoh:2002mn}, although the problem is essentially a five-dimensional one.  
For the strong-gravity regime in the RS models, the low energy (or derivative) expansion method would be a useful approach in some cases\cite{Wiseman:2002nn,Kanno:2002ia,Shiromizu:0210066}. 
A recent fully nonlinear numerical treatment of time-dependent braneworld models is also an interesting development to understand nonlinear brane dynamics \cite{Martin:2003yh}.

A pioneering approach to strongly gravitating systems in the RS model was developed by Wiseman in the study of relativistic stars on the brane \cite{Wiseman:2001xt}. 
In this study, the full five-dimensional Einstein equations are solved numerically, employing a conformal gauge method that reduces the Einstein equations to a suitable boundary value problem under an assumption of static axisymmetry. 
A consequence of nonlinear effects in standard four-dimensional Einstein gravity is the existence of an upper mass limit at a fixed radius of relativistic stars. 
The existence of the upper mass limits is confirmed for small stars whose proper radii ${\mathcal{R}}$ are less than the AdS curvature radius $\ell$.
While the upper mass limits for large stars whose radii are ${\mathcal{R}} \lesssim 3\ell$ are not confirmed, all observed physical quantities indicate that an effective four-dimensional description by means of four-dimensional Einstein gravity can be applied far into the nonlinear regime. 
Then the so-called recovery mechanism of Einstein gravity on the brane would persist in the strong-gravity regime.

The basic idea of the conformal gauge method is a key to exploring the nonlinear regime of gravity with extra dimensions. 
Indeed, it has been applied for nonuniform black strings \cite{Wiseman:2002zc,Wiseman:2002ti} and Kaluza-Klein black holes \cite{Kudoh:2003ki,Sorkin:2003ka}. 
As is emphasized in Ref. \cite{Wiseman:2002zc}, the problem of nonuniform black strings is a clean and simple application of the conformal gauge method compared with the other applications, due to a technical issue of numerics associated with the coordinate system at the symmetry axis (see, e.g., Refs. \cite{Alcubierre:1999ab,Shibata:2000} for a similar problem in numerical relativity, although the technical issue is intrinsic to the conformal gauge).

The method is also applied to a study of black holes in the RS braneworld model \cite{Kudoh:2003xz,Kudoh:2003vg}. 
Physically realistic black holes are thought to localize on the brane, as matter fields localize there \cite{Chamblin:1999by}. Static axisymmetric localized black holes, for example, would be formed as a result of gravitational collapse on the brane, settling down to a static state at late times.
There are many discussions about the localized black holes \cite{Chamblin:1999by,Chamblin:2000ra,Cadeau:2000tj,Casadio:2002uv,Casadio:2001jg,Dadhich:2000am,Garriga:1999bq,Giannakis:2000ss,Modgil:2001hm,Kanti:2001cj,Kofinas:2002gq,Kodama:2002kj,Shankaranarayanan:2003,Tamaki:2003bq,Vacaru:2001rf,Vacaru:2001wc} (recent discussions are \cite{Charmousis:2003wm,DeSmet:2003bt,Kanti:2003,Karasik:2003}). 
However, physically acceptable black hole solutions have not been found so far, except for exact solutions in the four-dimensional RS braneworld found by Emparan, Horowitz, and Myers (EHM)~\cite{Emparan:1999wa,Emparan:1999fd}.   
An interesting argument motivated by a no-go theorem \cite{Bruni:2001fd} and the AdS/CFT correspondence in the RS braneworld (e.g. Refs. \cite{Duff:2000mt,Hawking:2000kj}) is that large localized static black holes may not exist~\cite{Emparan:2002px,Tanaka:2002rb}.

In our previous papers, we considered a localized black holes in the five-dimensional RS single brane model, and presented some examples of numerical solutions \cite{Kudoh:2003vg,Kudoh:2003xz}.  
However, we could not succeed in constructing localized black holes that are larger than the AdS radius, because the numerical scheme gave no convergent solution for them.
To overcome the difficulty and to obtain an understanding of localized black holes, we need further development of the method. A key method had been developed through the study of Kaluza-Klein black holes \cite{Kudoh:2003ki} (see also Ref. \cite{Wiseman:2001xt}).  In this paper, we apply the improved numerics and discuss numerical solutions of static localized black holes.   
We perform the analysis in the six-dimensional RS model. This is a technical reason that numerical calculations in higher dimensions have an advantage: The metric in higher dimensions decay faster than that in the original five-dimensional braneworld model, and which makes numerical calculations more tractable \cite{Wiseman:2002zc}. 
We think that there is no qualitative difference between localized black holes in five-dimensions and those in six-dimensions. 
Indeed, as we will see, thermodynamic properties of the small localized black holes in five-dimensions found in Refs. \cite{Kudoh:2003vg,Kudoh:2003xz} persist in six-dimensions.  
While the EHM solutions are known in four-dimensions, the difference between four-dimensions and dimensions more than four is very large.
For example, the EHM black holes have a maximum mass, corresponding to the asymptotic deficit angle on the brane.

We are interesting in qualitative properties of the localized black holes, for example, the thermodynamic property, horizon geometry and gravity induced on the brane. 
In particular, it is an interesting question whether the lower-dimensional Einstein gravity is recovered on the brane even in the presence of a localized black hole.  
We construct the static localized black holes by means of the numerical method and discuss the fundamental properties of the black holes.

In Section  \ref{sec:6D braneworld}, we begin by briefly reviewing the conformal gauge method, and then in Section  \ref{sec:Numerical solutions}, we present numerical solutions, showing the performance of the method. 
Properties of the localized black holes and their implications are discussed in Section  \ref{sec:Properties of LBH}.  Section  \ref{eq:conclusion} is devoted to a summary. 
Due to the technical nature of Section \ref{sec:Numerical solutions}, the reader may prefer to skip straight to Section \ref{sec:Properties of LBH}.

\section{Six-dimensional braneworld black holes}
\label{sec:6D braneworld}

\subsection{Conformal gauge method}
\label{subsec:Conformal gauge method}

Following previous work \cite{Kudoh:2003xz,Kudoh:2003vg}, we adopt the conformal gauge method to construct localized black holes.
We begin with a general ansatz for static axisymmetric black hole solutions to Einstein's equations with a bulk cosmological constant in $D$ dimensions: 
\begin{eqnarray}
ds^2 = \frac{\ell^2}{z^2} 
\left(
  - e^{2\alpha} dt^2 +e^{2 \beta } (dr^2+dz^2)+r^2 e^{2\gamma} d\Omega^2_{D-3}
\right) \,,
\label{eq:assume RS BH}
\end{eqnarray}
where the metric functions are arbitrary functions of $r$ and $z$. 
The bulk cosmological constant is $ \Lambda = -  {(D-1)(D-2)}/{2\ell^2 }$, and tension of the branes is 
$\sigma = 2(D-2)/\kappa_D \ell$. 
Here $\kappa_D=8\pi G_D$ is $D$-dimensional Newton's constant and $\ell$ is the AdS curvature radius.  Israel's junction condition at the brane in $D$-dimensions gives 
$
     K_{\mu\nu}|_{z= \ell + } 
   = - {\gamma_{\mu\nu}}/{ \ell } \,,
$
where  $K_{\mu\nu}$ is the extrinsic curvature of the brane at $z= \ell$ with the induced metric $\gamma_{\mu\nu}$.

Now we consider a six-dimensional (6D) single brane model. 
The metric form has residual gauge degrees of freedom of conformal transformations in the two-dimensional space $\{r,z \}$, and then we can use these conformal degrees of freedom to transform the location of the event horizon to be at a constant radius 
\begin{eqnarray}
\rho =  \rho_0. 
\end{eqnarray}
Here we have introduced polar coordinates, 
\begin{eqnarray}
&& r = \rho \sin \chi,
\nonumber
\\
&& z = \ell + \rho \cos \chi \,.
\end{eqnarray} 
Note that the positive tension brane can be placed at $z=\ell$.  
We introduce 
\begin{eqnarray}
    \xi = \chi^2. 
\label{eq: xi}
\end{eqnarray}
This variable is used in Ref. \cite{Kudoh:2003xz} to gain numerical stability near the axis. 
However, even with this modification, the algorithm is still unstable and encounters a lack of convergence when we increase the radius of black hole. 
As we discuss later, this numerical problem is resolved in Ref. \cite{Kudoh:2003ki}, altering a method originally used in Ref. \cite{Wiseman:2001xt}. 
An aim in using the variable (\ref{eq: xi}) is to increase a numerical resolution near the brane. 
All numerical calculations presented in this paper are performed by using the variable (\ref{eq: xi})    
\footnote{
The introduction of the variable $\xi$ is not essential in our numerics, but the increased resolution makes numerics work better, compared with calculations using the variable $\chi$.  Indeed, this allows us to find larger black holes.  
}.

If the localized black hole is sufficiently small compared to the curvature radius in the bulk, the black hole would be approximated by the 6D Schwarzschild black hole because the tension of the brane and the bulk cosmological constant would not affect locally the black hole. This is in fact observed in the study of 5D braneworld black holes \cite{Kudoh:2003vg}.  
Thus we subtract the 6D Schwarzschild metric from the assumed metric.
Rewriting the metric functions as \cite{Kudoh:2003xz}
\begin{eqnarray}
\alpha &=& A_0 ,
\cr
\beta &=& B_0 - C_0 ,
\cr
\gamma &=& B_0 + \frac{2}{3} C_0 ,
\end{eqnarray}
and we decompose these into
\begin{eqnarray}
  X_0(\rho, \xi) = X_S(\rho) + X (\rho, \xi ) .
\end{eqnarray}
where $X = A, B$, or $C$. Each $X_S$ is defined by
\begin{eqnarray}
A_S(\rho) &=& \log \left( \frac{ \rho^3 - \rho_0^3 }{ \rho^3 + \rho_0^3 } \right) \,,
\cr
B_S(\rho) &=& \frac{2}{3} \log \left( 1 +\frac{\rho_0^3}{\rho^3 } \right)\,,
\cr
C_S (\rho) &=&0 \, .
\end{eqnarray}
When we take $A=B=C=0$ and $\ell\to\infty$ ($\ell/z \to 1$), the 6D Schwarzschild solution in isotropic coordinates is recovered.

Black hole solutions are parametrized only by a parameter
\footnote{
\label{footnote:1}  
In our numerical calculations, we take $\rho_0=1$ and change $\ell$ to specify $\hat{\rho}_0$  \cite{Kudoh:2003xz}.
Because the variation of $\ell$ keeping $\hatrho$ fixed corresponds to rescaling the length scale, we must be careful to compare physical quantities of a black hole to those of a black hole with different $\hatrho$. 
Thus in this paper we plot physical quantities in dimensionless form. 
We also note that physical size of the grid resolution ($d\xi$, $d\rho$) changes for each calculation due to the rescaling of the length scale, even if we fix the grid resolution. 
}
\begin{eqnarray} 
\hatrho = \frac{\rho_0}{\ell} \,. 
\end{eqnarray}
As we will see later (Fig. \ref{fig:aspectratio}), the proper circumferential radius of the horizon on the brane is approximately equal to the parameter $\hatrho$. 
Black holes with $\hatrho < 1$ are small compared with the AdS curvature radius, whereas black holes with $\hatrho > 1$ are large.

\subsection{Einstein equations} 

The Einstein equations yield five nontrivial equations. Three of them yield elliptic equations for the respective metric components $X=\{A, B, C \}$, 
\begin{eqnarray}
    (\partial_r^2 + \partial_z^2) X = S_{X} ,
\label{eq:elliptic eqs}
\end{eqnarray}
where the source $S_{X}$ depends nonlinearly on $X$ and its derivatives. 
The remaining two equations give constraint equations, which satisfy Cauchy-Riemann relations. The Cauchy-Riemann relations are obtained from the nontrivial components of the Bianchi identities $\nabla^\mu G_{\mu \nu}=0$, assuming that the equations for $A$, $B$, and $C$ are satisfied.  The Cauchy-Riemann relations in terms of $\{ \rho,\xi \}$ are 
\begin{eqnarray}
&&   \frac{2\sqrt{\xi}}{\rho} \partial_\xi \Phi +\partial_\rho \Psi =0 \,,
\\
&&   \frac{2\sqrt{\xi}}{\rho}\partial_\xi \Psi -\partial_\rho \Phi =0 \,,
\end{eqnarray}
where 
$\Phi:= 2\xi \sqrt{-\mathrm{det} g }\, { G}^\rho_\xi $,
and 
$\Psi:= \rho \sqrt{\xi} \sqrt{-\mathrm{det} g}({G}^\rho_\rho -{G}^\xi_\xi). $  
$\Psi$ and $\Phi$ are called weighted constraint equations. 
Since $\Psi$ and $\Phi$ satisfy Cauchy-Riemann relations, each of them satisfies the Laplace equation. 
Hence if $\Psi$ is satisfied on all boundaries and provided that $\Phi$ vanishes at any one point, the two weighted constraints must vanish in all places as long as the elliptic equations for $A, B$, and $C$ are satisfied.  

\subsection{Boundary conditions} 

To solve the elliptic equations (\ref{eq:elliptic eqs}) numerically, we need boundary conditions for them. 
The boundary conditions are determined by physical requirements. 
Let us first consider the boundary conditions on the event horizon.
The horizon is a regular singular point of the elliptic equations for fixed $\chi$. The regularity at the horizon demands that singular solutions should be absent. Thus a power expansion in $\rho$ of the elliptic equations and the constraint equation $(G^{\rho}_{\rho} - G^\xi_\xi)=0$ results in boundary conditions at the horizon for regularity:
\begin{eqnarray}
\partial_\rho A = \partial_\rho  B = \frac{\cos\chi}{z}\,, 
\quad
\partial_\rho  C = 0. \quad ( \rho=\rho_0)
\end{eqnarray}
In addition to these conditions, the regularity of the constraint equation $G^{\rho}_{\xi}=0$ gives a condition 
\begin{eqnarray}
    \partial_\chi(A- B +C )=0. \quad ( \rho=\rho_0)
\label{eq: T=const.}
\end{eqnarray}
This additional requirement is physically a condition that temperature of a black hole is constant on the horizon (see Eq. (\ref{eq:T S})). 
We do not use this boundary condition for the elliptic equations, but we use it to integrate the constraint equation $\Phi$ (see Sec. \ref{subsec:stabilization}).

Boundary conditions on the brane are derived from the Israel's junction condition. They are given by
\begin{eqnarray}
&& 
\partial_{\chi}  A  = \partial_{\chi} B 
  =  - \frac{\rho}{ \ell} 
  \left[
   1-  \left(1 +\frac{\rho_0^3}{\rho^3} \right) e^{B-C}
   \right] 
  \,,\quad  (  \chi=  {\pi}/{2}  ).
\cr
&& \partial_{\chi} C  =0.
\end{eqnarray}

On the axis $r=0$, the absence of conical singularity requires 
\begin{eqnarray}
    C(\rho, 0)=0. 
   \label{eq: C=0}
\end{eqnarray}
This condition is also required from the regularity of the equation for $C$ on the axis. 
Furthermore, the axial symmetry requires that the $r$ derivative of the metric functions vanishes on the axis. 
Thus we have 
\begin{eqnarray}
    \partial_\chi A, ~~ \partial_\chi B,~~  \partial_\chi C =0.  \quad(\chi=0)
\label{eq: ABC on axis }
\end{eqnarray}

Let us consider the asymptotic boundary conditions. 
The background spacetime of the RS braneworld is the AdS spacetime. 
It is well known that localized matter fields reproduce linearized lower-dimensional Einstein gravity on the brane in the weak gravitational regime, and the linear perturbations fall off asymptotically keeping the spacetime to be asymptotically AdS. Thus in the current problem, we could assume that the spacetime is asymptotically AdS. 
Therefore, the asymptotic boundary conditions are 
\begin{eqnarray}
    A, B, C = 0  ~~(\rho \to \infty)
\label{eq: ABC at infty}
\end{eqnarray}
at spatial infinity, although in actual numerics these conditions are imposed at a finite coordinate location.  
In Appendix \ref{sec:asymptotic bc}, we discuss asymptotic behavior of gravitational field induced by a point mass on the brane.  
The gravitational field in the asymptotic region is anisotropic in the polar coordinates \cite{Garriga:1999yh,Sasaki:1999mi,Giddings:2000mu}.
While the asymptotic behavior given by the point mass could provide asymptotic boundary conditions for the localized black holes as Neumann boundary conditions, we have not been able to find a stable scheme involving the boundary conditions. However, already with our simple implementation it is possible to derive interesting results as we will see (see Sec. \ref{sec:Performance of the method} for tests of our numerical method)

\section{Numerical solutions}
\label{sec:Numerical solutions}

\begin{figure}
\begin{center}
  \includegraphics[width=7cm,clip]{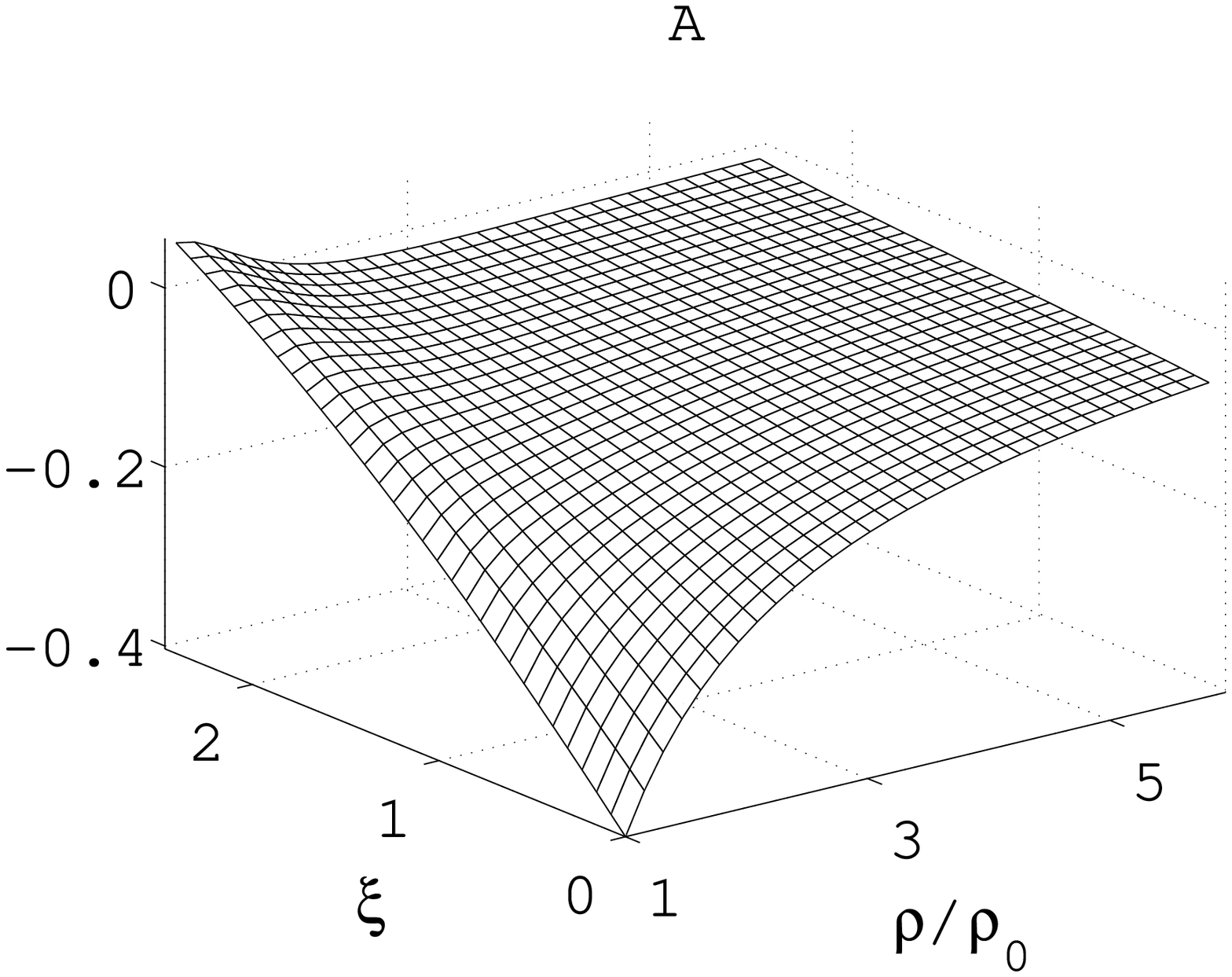}
  \includegraphics[width=7cm,clip]{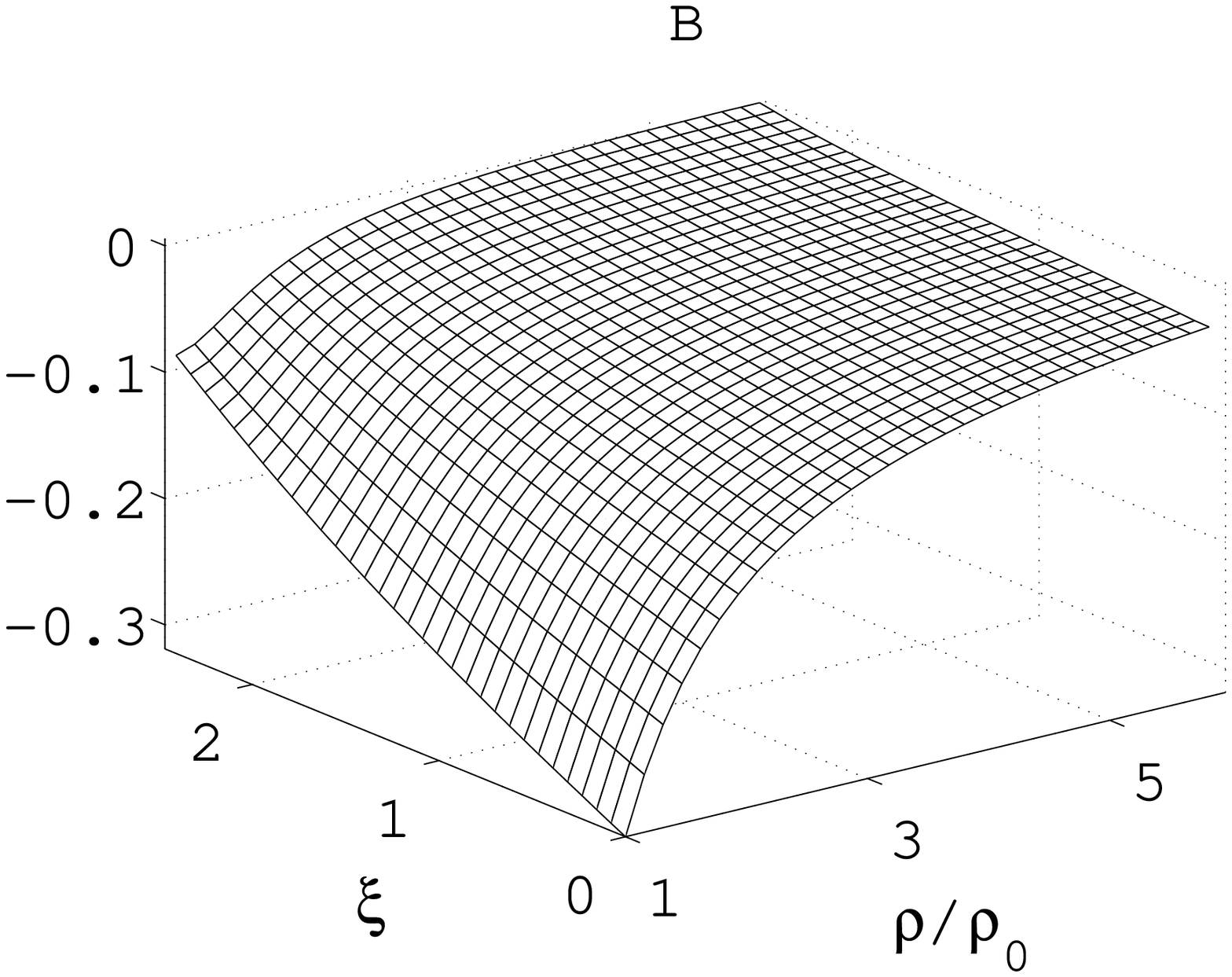}
\end{center}
\begin{center}
  \includegraphics[width=7cm,clip]{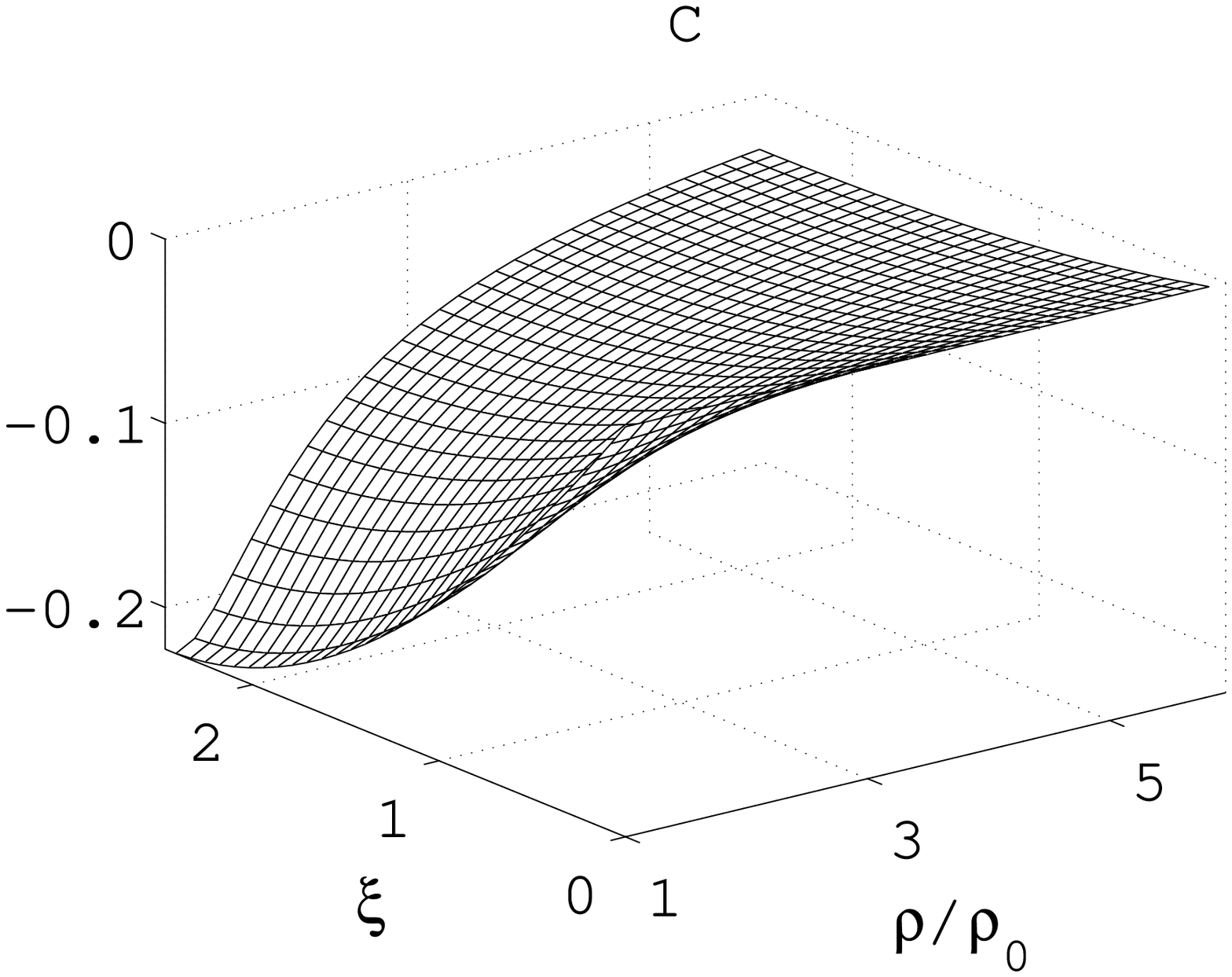}
  \includegraphics[width=7cm,clip]{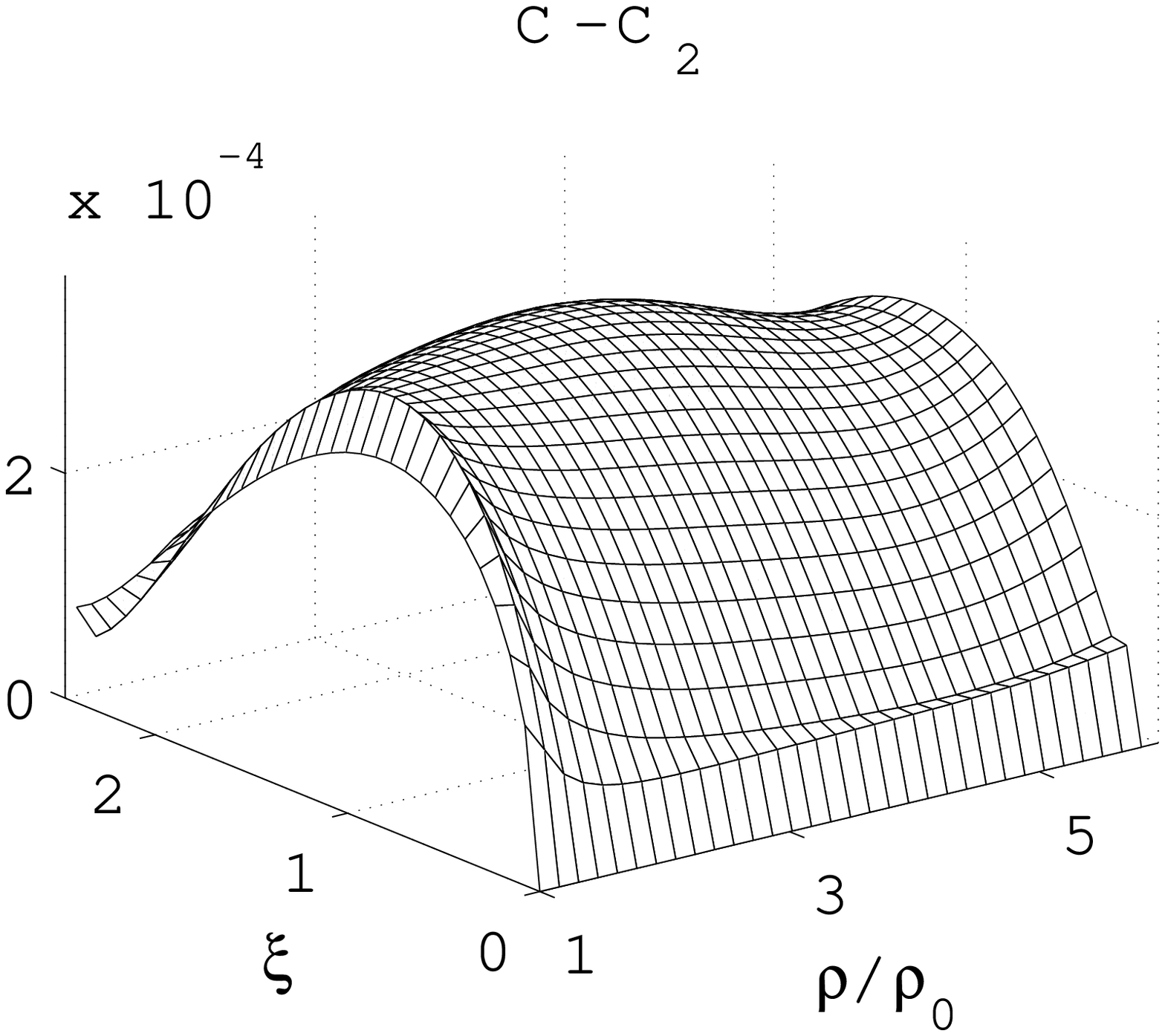}
\end{center}
\caption{ 
\label{fig:ABC L2}
Metric functions and $C-C_2$ for $\hatrho=0.5$. 
}
\end{figure}

\begin{figure}
\begin{center}
  \includegraphics[width=7cm,clip]{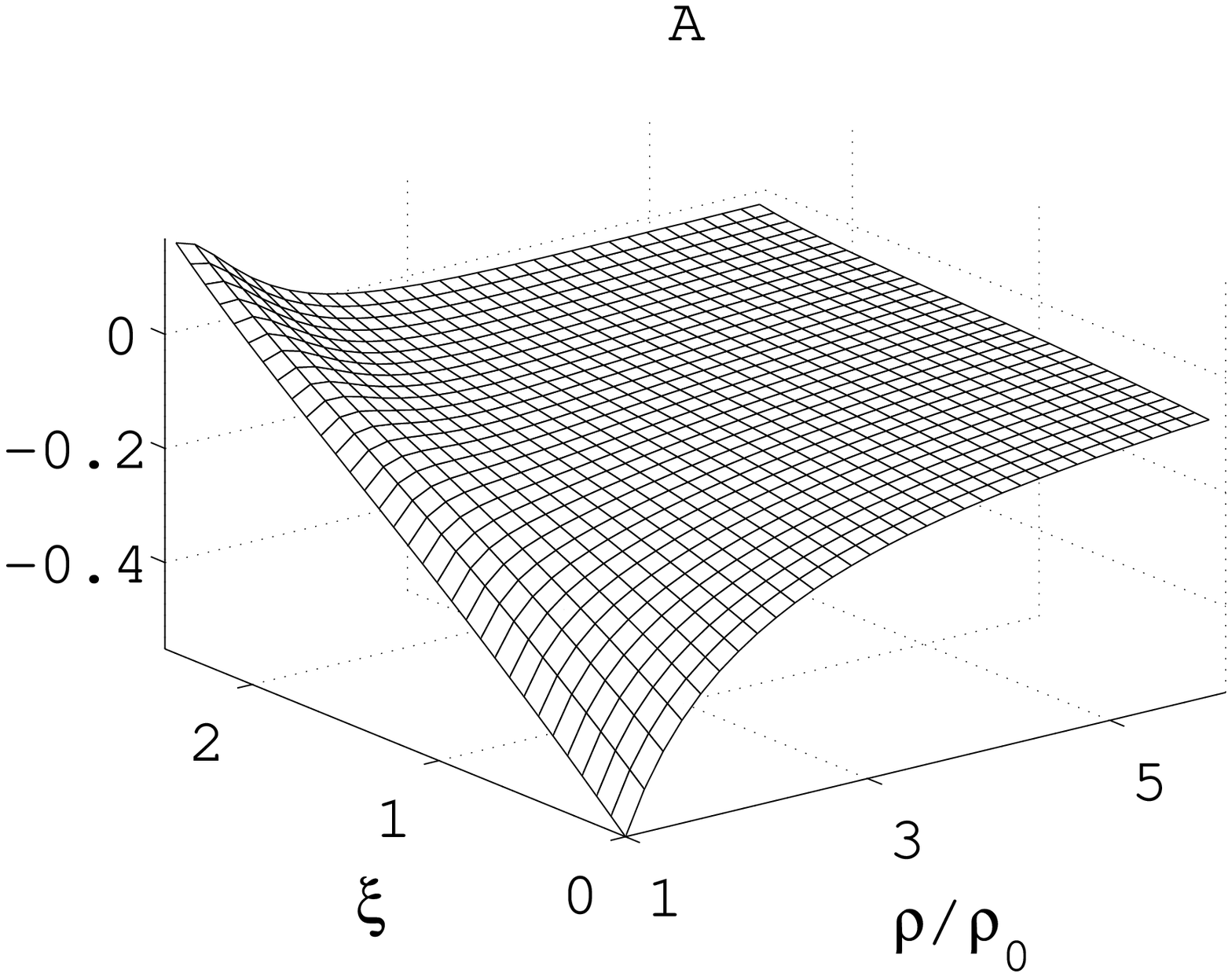}
  \includegraphics[width=7cm,clip]{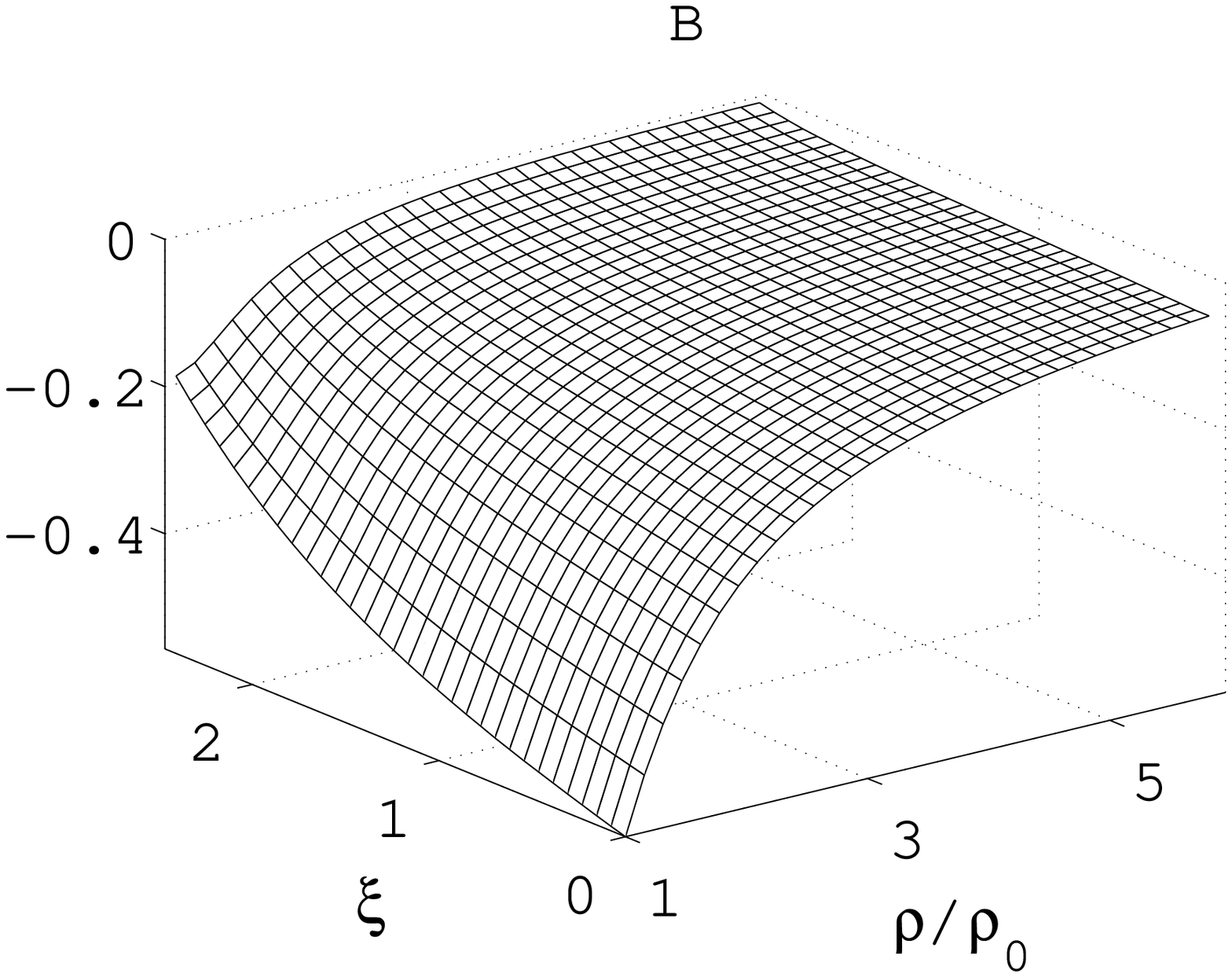}
\end{center}
\begin{center}
  \includegraphics[width=7cm,clip]{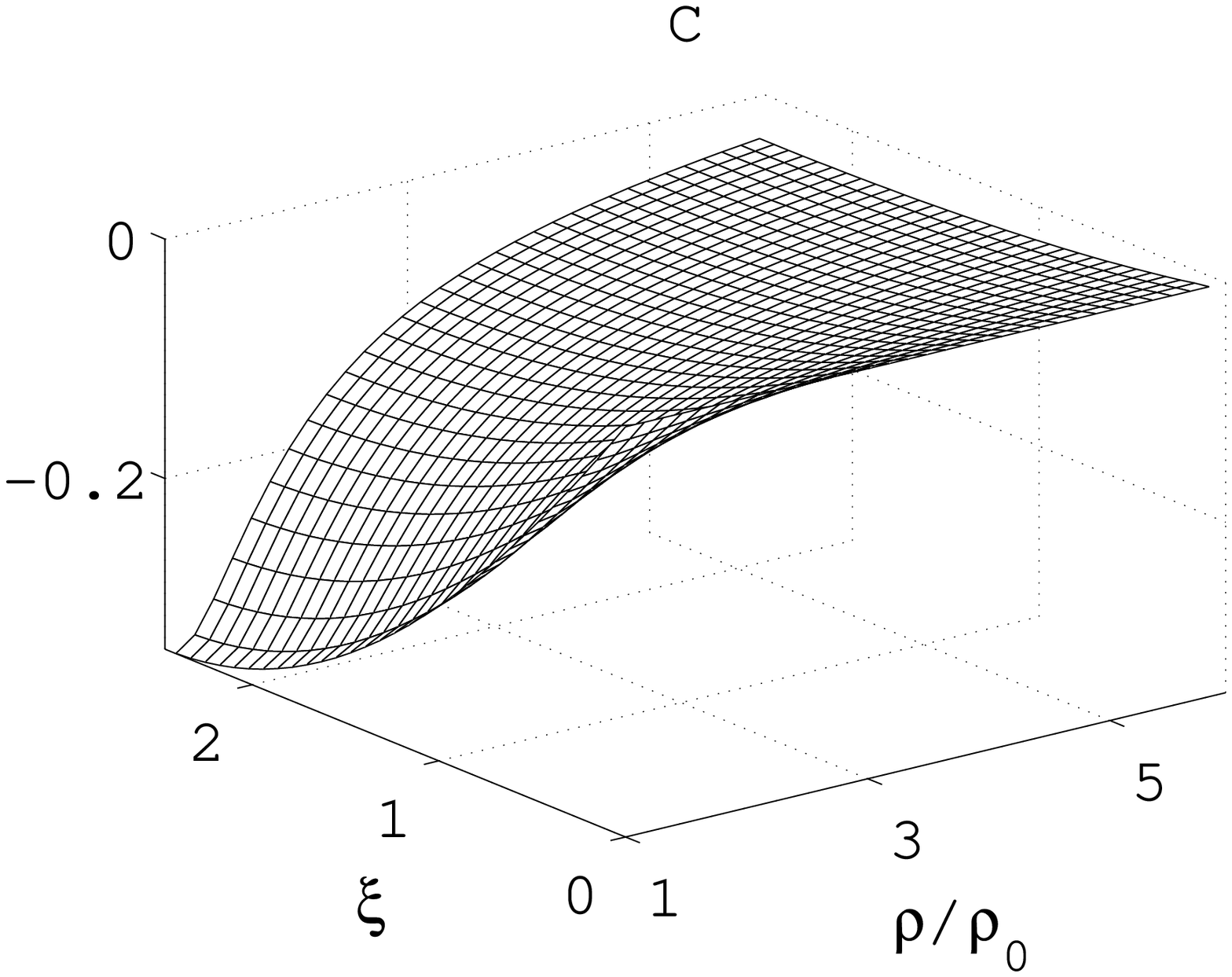}
  \includegraphics[width=7cm,clip]{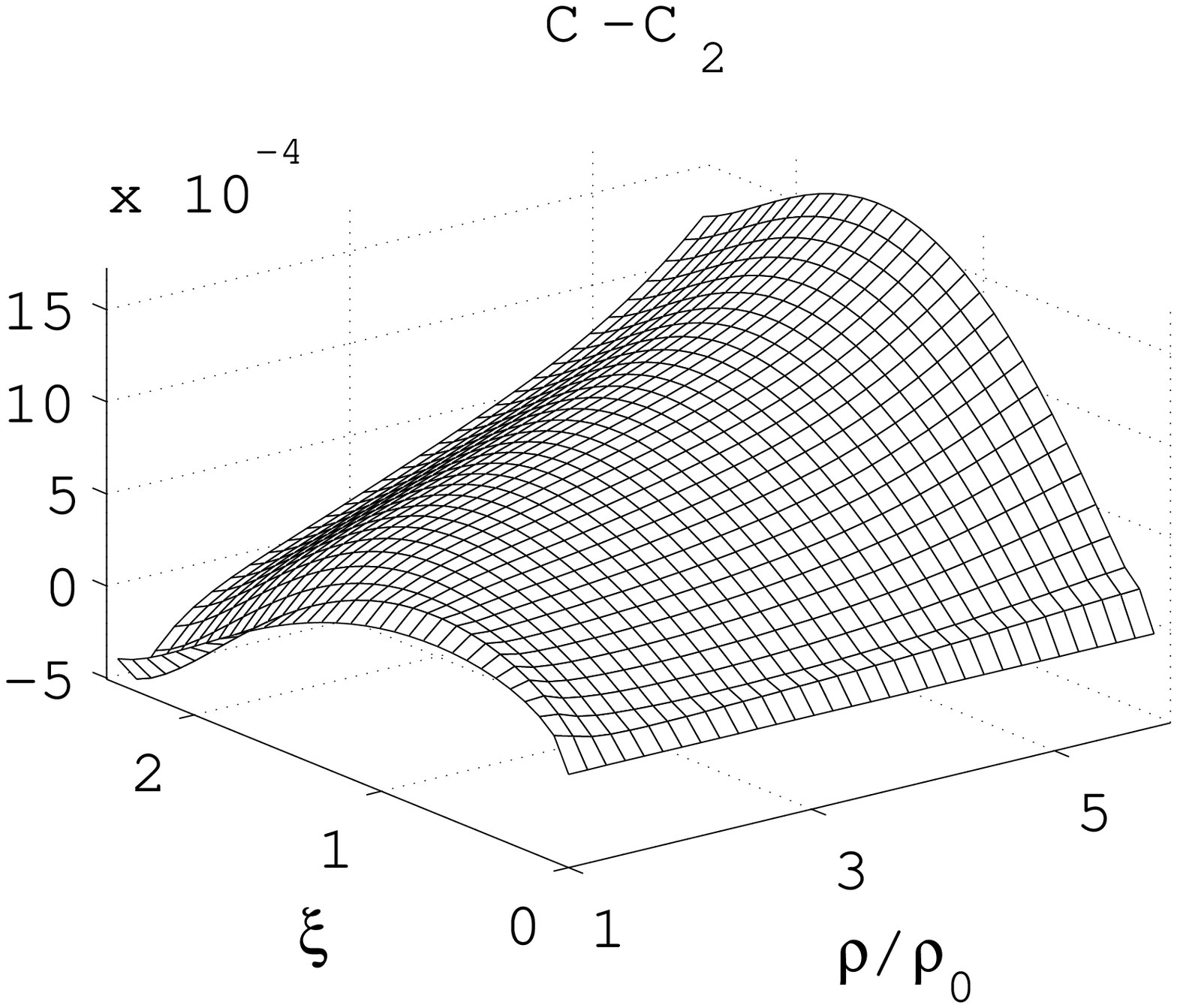}
\end{center}
\caption{ 
\label{fig:ABC L08}
Metric functions and $C-C_2$ for $\hatrho=1.25$. 
}
\end{figure}

\begin{figure} 
\centerline{
  \includegraphics[width=7cm,clip]{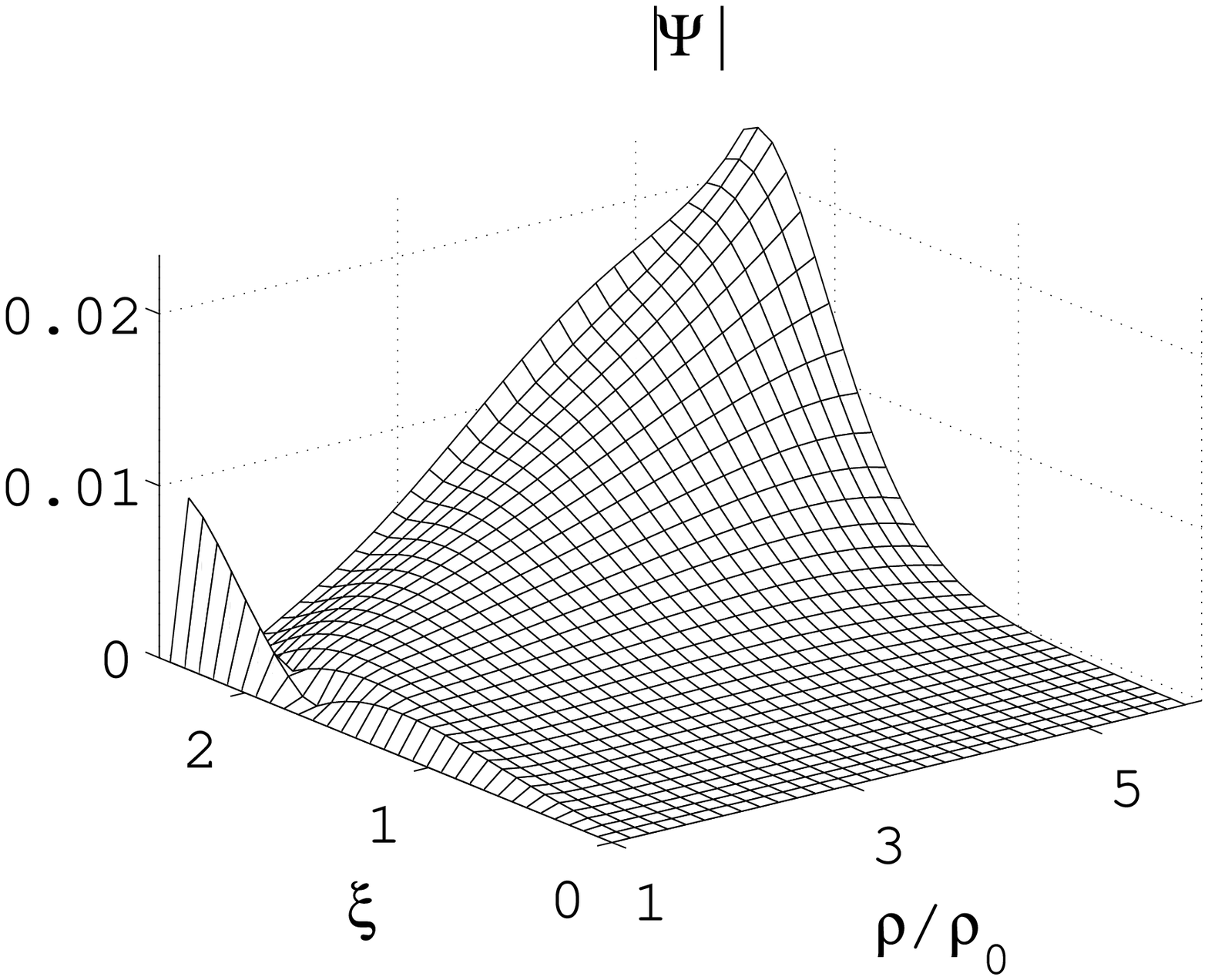}
  \includegraphics[width=7cm,clip]{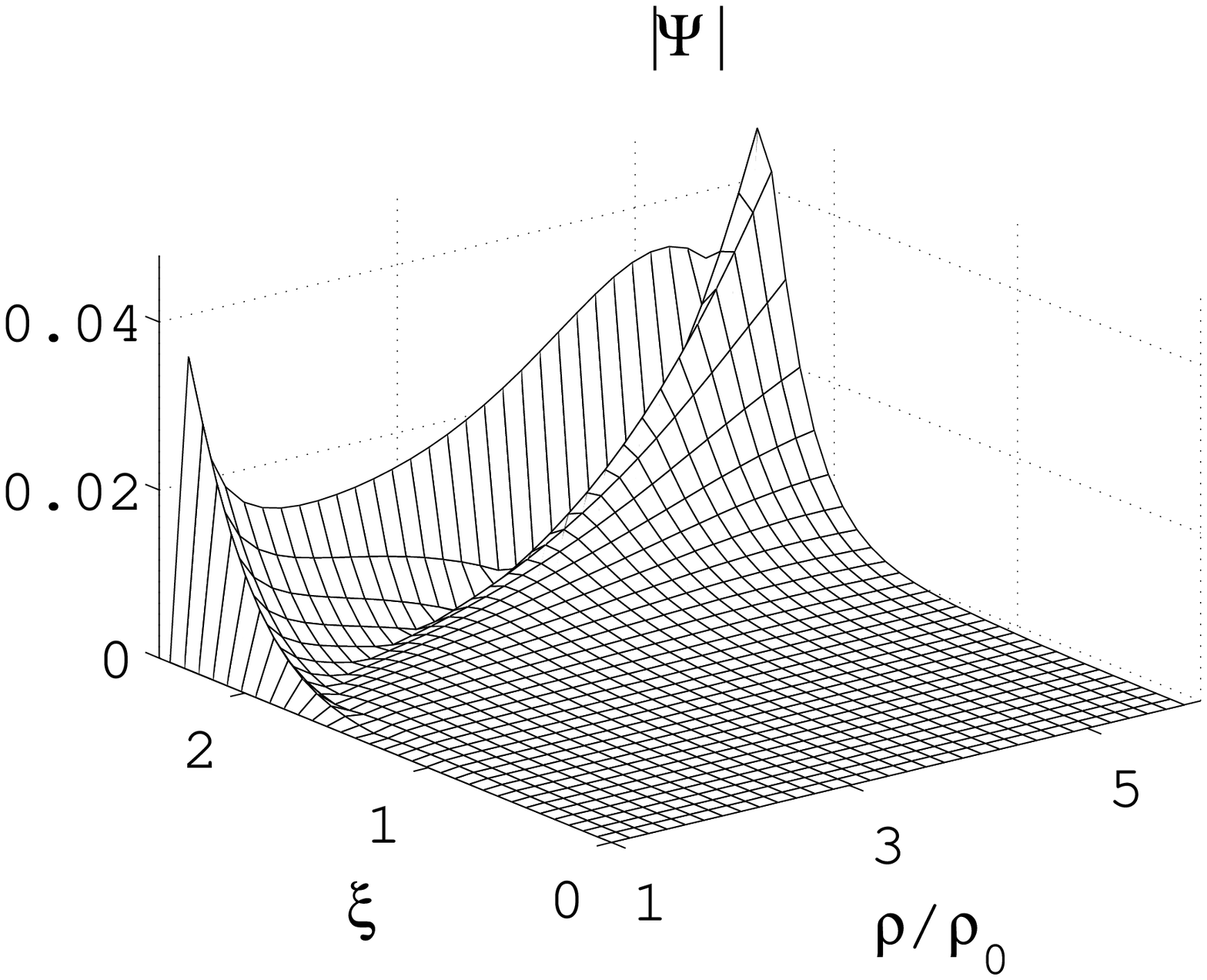}
}
\caption{ 
\label{fig:constraints L2}
Absolute values of the weighted constraint equation $\Psi$ for $\hatrho=0.5$ (left) and $\hatrho=1.25$ (right). 
}
\end{figure}

\begin{table}[t]
\begin{center}
\begin{tabular}{c |cc|cc| cc }
\hline   \hline
           & $  \left \langle |\Psi| \right\rangle $ &
           & $  \left \langle |\Phi| \right\rangle $ & 
           & $  \left \langle |C-C_2| \right\rangle $ & 
\\   
& $\hatrho=0.5 $ &1.25  
& 0.5 & 1.25
& 0.5 &1.25 
\\   \hline
$32\times 128$
&4.1 $\times 10^{-2}$  &3.4 $\times 10^{-1}$ 
&1.9 $\times 10^{-2}$  &4.0 $\times 10^{-2}$ 
&3.1 $\times 10^{-3}$  &7.1 $\times 10^{-3}$   
\\   \hline
$64 \times 256$
&1.1 $\times 10^{-2}$ &7.8 $\times 10^{-2}$ 
&4.9 $\times 10^{-3}$  &1.6 $\times 10^{-2}$  
&8.7 $\times 10^{-4}$  &2.0 $\times 10^{-3}$  
\\   \hline
$128 \times 512 $
&3.5 $\times 10^{-3}$  &1.4 $\times 10^{-2}$   
&1.6 $\times 10^{-3}$  &3.7 $\times 10^{-3}$  
&2.5 $\times 10^{-4}$  &5.5 $\times 10^{-4}$  
\\
\hline \hline   
\end{tabular}
\caption[short]{
Averaged violations of the absolute values of the weighted constraints and $C-C_2$ for the three resolutions. 
The three resolutions converge consistently with a second-order scaling.
The averaged violations are calculated for a fixed domain $1<\rho/\rho_0<5$. 
}  
\label{table:3 resolutions}
\end{center}
\end{table}

\begin{table}[t]
\begin{center}
\begin{tabular}{r|cc|cc| cc |cc|cc}
\hline   \hline
           & $ \delta \left \langle |\Psi| \right\rangle $  &
           & $ \delta \left \langle |\Phi| \right\rangle $ & 
           & $  {\mathcal{\delta T} }  $  &  
           & $  {\mathcal{\delta S} }_6$  &
           & $  {\mathcal{\delta S} }_5$  &
\\   
& $\hatrho=$0.5, &1.25  & 0.5, & 1.25
& 0.5, &1.25   & 0.5, &1.25
& 0.5, &1.25
\\   \hline
$\rho_{max}/\rho_0=$ 6.0
&$~$0.673 & $~$0.369 &  $~$12.9$~$  & $~$1.32$~$
&$-$0.014 & $-$0.029 & $~$0.082 & $~$0.133 
&$~$0.071 & $~$0.126 
\\   \hline
7.2
&$-$0.020 & 0.319   & 3.12   & 0.594
&$-$0.009 & $-$0.017 & 0.049  & 0.081 
&$~$0.043 & $~$0.077
\\   \hline
9.7
&$~$0.016 & 0.145 &$-$0.104 & 0.164
&$-$0.003 & $-$0.006 & 0.016 & 0.027 
&$~$0.014 & $~$0.026
\\   \hline
11.0
&$~$0.011 & 0.060 &$~$0.000 & 0.155
&$-$0.001 & $-0.003$ & 0.007 & 0.012
&$~$0.006 & $~$0.011
\\
\hline \hline   
\end{tabular}
\caption[short]{
Sensitivity of the solutions to the finite position of the asymptotic boundary $\rho_{max}$. 
We compare the temperature, entropy, and weighted constraints for solutions with $\hatrho =0.5$ and  $1.25$. 
 The variation of the quantities is defined as $  {\mathcal{\delta T} }= {\mathcal{ T}}/{\mathcal{ T}}_{(\rho_{max}=12.2)} - 1 $, dividing by ${\mathcal T}$ for ${\rho_{max}}=12.2\rho_0$.  
The averaged absolute values of the weighted constraints, $\left \langle |\Psi| \right\rangle $ and $\left \langle |\Phi| \right\rangle $, are calculated for a fixed domain, $1<\rho/\rho_0<5$.
The constraints appear to be even better satisfied as $\rho_{max}$ increases. Moreover, thermodynamic quantities hardly change, indicating our choice of $\rho_{max}$ is sufficiently large.  
All solutions are at the same resolution as for $64\times 256$ with $\rho_{max}=11\rho_0$.
}
\label{table:rho_max}
\end{center}
\end{table}

\subsection{Stabilization of numerical scheme}
\label{subsec:stabilization}

In this section, we discuss the stability of our relaxation scheme of numerics.  
A problem of solving the elliptic equations (\ref{eq:elliptic eqs}) by a relaxation scheme is an instability coming from the axis ($r=0$). A term in the equation for $C$ that has a factor $1/r^2$ gives severe contribution near the axis, and it invokes numerical instability~\cite{Kudoh:2003xz}. Actually if we turn off the term during a calculation, then the calculation becomes very stable and converges to a solution, although the solution is not a solution of Einstein's equations. 
This issue of instability prevents us from finding large localized black holes in our previous work.

We deal with this term as in Ref. \cite{Kudoh:2003ki}. A second derivative term for $C$ in the constraint equation $\Phi \propto {G}^\rho_\xi$ is simply $C_{,\rho\xi}$, and thus we can integrate it over the domain.  
A point is that this equation has no singular term as long as Eq. (\ref{eq: ABC on axis }) is satisfied. Thus the solution for $C$ integrated away from the axis has very good behavior near the axis. We call the solution $C_2$ to distinguish it from $C$. 
Given the characteristics we need two initial data surfaces, one at constant $\xi$, and the other at constant $\rho$. 
Suitable surfaces are $\xi=0$ and $\rho=\rho_0$. 
For $\xi =0$ we can set $C_2 = 0$ to satisfy the condition (\ref{eq: C=0}), and this ensures the good quality of the $C_2$ behavior near the axis. 
For the other initial surface $\rho=\rho_0$, we fix $C_2$ using the condition (\ref{eq: T=const.}) that the horizon temperature is a constant.
We then integrate $C_2$ from these two boundaries: 
\begin{eqnarray}
C_2 (\rho, \xi) &=&  \int^\xi_0 d\xi  \int^\rho_{\rho_0} d\rho ~F
       + C(\rho_0, \xi) \,,
\cr
    C(\rho_0,\xi) &=&  (B-A) - (B-A)|_{\xi=0} \,. 
\label{eq:C2}
\end{eqnarray}
where $F$ is the remaining terms in the constraint equation. 
We calculate the $1/r^2$ term in the elliptic equation for $C$ using this function $C_2$, rather than $C$.  
Then the term in question is replaced by 
\begin{eqnarray}
 \frac{1}{r^2} \left( 1- e^{\frac{10}{3}C}\right)
\to
 \frac{1}{r^2} \left( 1- e^{\frac{10}{3}C_2}\right) \,. 
\end{eqnarray} 
This prescription dramatically improves stability of numerics, and we are free of the instability coming from the axis. 
Furthermore, we use condition (\ref{eq: T=const.}) in our scheme, the variation of the horizon temperature is vanishingly small $(< 10^{-2} \% )$, and the zeroth law of black hole thermodynamics is well satisfied.

\subsection{Performance of the method}
\label{sec:Performance of the method}

The formulations and numerical scheme described in the previous sections allow us to solve Einstein's equations as a boundary value problem.
To solve the problem numerically, we use a relaxation method with a second-order discretization scheme.  In this section, we present some results of numerical calculations and discuss qualitative aspect of numerical solutions.

An illustration of metric functions for $\hatrho=0.5$ and $\hatrho=1.25$ are shown in Figs. \ref{fig:ABC L2} and \ref{fig:ABC L08}. The former case is a typical for small localized black holes ($\hatrho<1$) and the latter case is for large localized black holes ($\hatrho>1$).
As we see, the global behavior of metric functions hardly change for the two cases, but the peak values of metric functions increase, indicating that nonlinear effects become much more important for large black holes. It is remarkable that $C-C_2$ is very small, and then the integration method works consistently. 
We have performed the calculations for three different resolutions, $32\times 128$,  $64 \times 256$, and $128\times 512$, setting $d\xi= d\rho$ (see Table \ref{table:3 resolutions}). 
The finite position of the asymptotic boundary is at 
\begin{eqnarray}
 \rho_{max} = 11 \rho_{0} .
\label{eq:rhomax}
\end{eqnarray}

In our simple implementation, largest horizon radius of a black hole that we could find depends on the resolution of the calculation, and then it is limited by computation time. 
The largest black hole that we found within a moderate computational time is $\hatrho = \rhomax$ with the grid resolution $128\times 512$.  
The localized black holes are systematically constructed for the parameter region $\hatrho = 0.002 \sim \rhomax $.

Figure \ref{fig:constraints L2}  shows absolute errors of the weighted constraint equation $\Psi \propto (G^{\rho}_{\rho} -G^{\xi}_{\xi}) $.  
Because $\Phi \propto G^\rho_\xi$ is essentially the same as $\Psi$ due to the Cauchy-Riemann relations and we have confirmed the consistency of $G^\rho_\xi$ by showing $C-C_2$ in Figs. \ref{fig:ABC L2} and \ref{fig:ABC L08}, we suppress figures of $\Phi$. 
While the corresponding Einstein equation $G^{\rho}_{\rho} -G^{\xi}_{\xi}=0$ is well satisfied yielding vanishingly small values, peaks of the errors of the weighted constraints appear in the asymptotic region near the brane. 
The peaks are originated from the huge amplification of the tiny violations of Einstein's equations due to the weight function $\rho \sqrt{-g}$ that is used to define $\Psi$.  
The weight function is proportional to $\sim \rho^5$ near the brane.  
Because the tiny violation of Einstein's equations comes from the asymptotic boundary, we will need to avoid using the naive asymptotic boundary conditions (\ref{eq: ABC at infty}).
As we have discussed, the naive Dirichlet conditions could be replaced by appropriate Neumann conditions.
The prescription will reduce such errors from finite boundary effects, if we can implement it without any numerical instability. 
Nevertheless, as $\rho_{max}$ increases, the approximation using the Dirichlet boundary conditions (\ref{eq: ABC at infty}) becomes much better, and then violations of the weighted constraints decrease.  
This is observed in our numerics (Table \ref{table:rho_max}). The averaged absolute errors of the weighted constraints in a fixed domain, $\left\langle|\Phi|\right\rangle$ and $\left\langle|\Psi|\right\rangle$, decrease as $\rho_{max}$ increases. Thus, we do not care about the violations of the weighted constraints in the asymptotic region in the present numerics.  
In Table \ref{table:rho_max}, we also show that errors from finite boundary effects in thermodynamic quantities are $\lesssim 1\%$ for 
Eq. (\ref{eq:rhomax}), and then the location of the boundary is sufficiently far.

\section{Properties of the localized black holes }
\label{sec:Properties of LBH} 

\subsection{Thermodynamic quantities}

\begin{figure}
\begin{center}
 \includegraphics[width=8cm,clip]{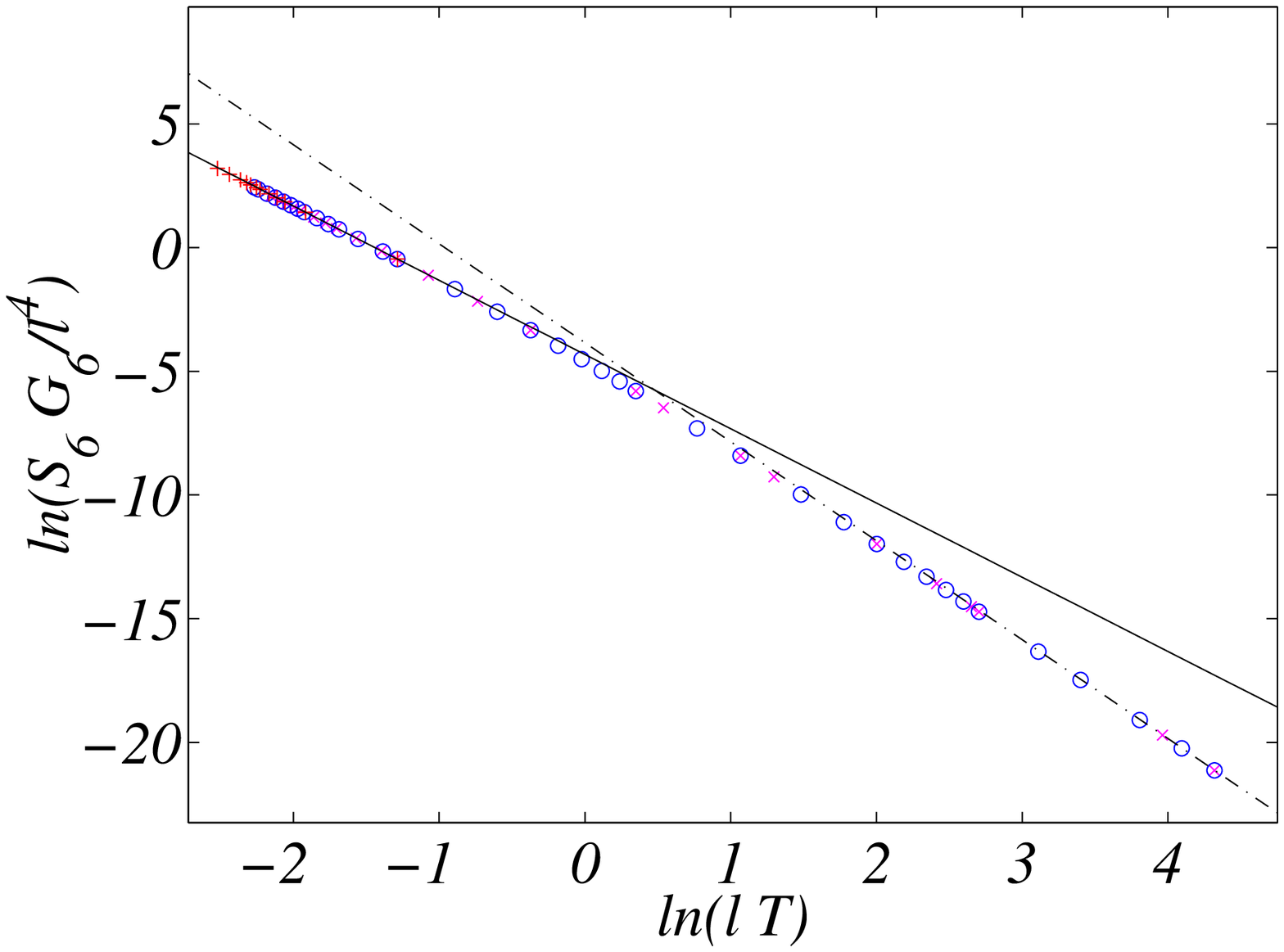}
 \quad
 \includegraphics[width=8cm,clip]{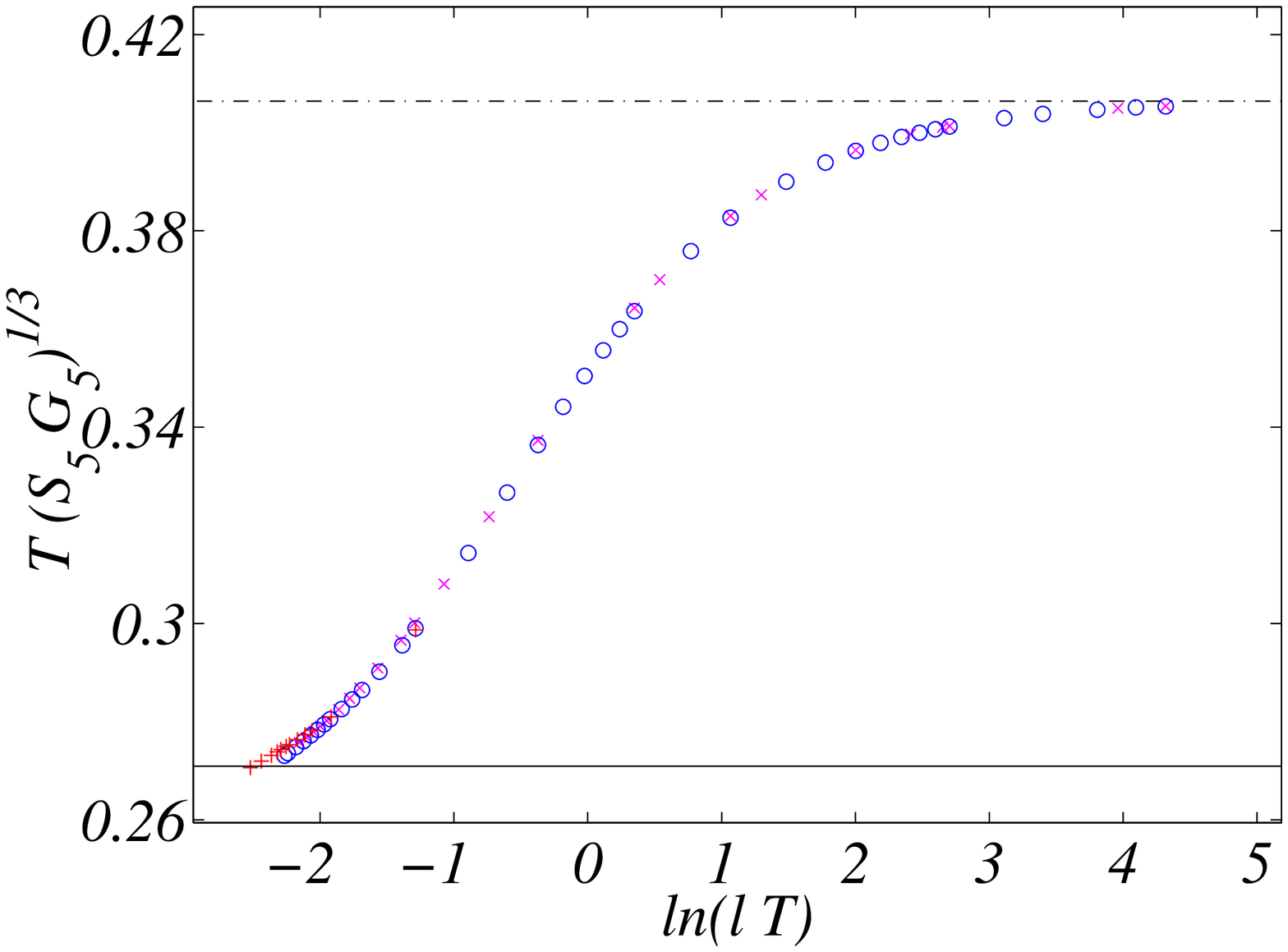}
\end{center}
\caption{ 
\label{fig:S5T and S6T}
Thermodynamic quantities of the localized black holes.
The left figure shows a thermodynamic relation for the temperature $\mathcal{T}$ and the entropy ${\mathcal S}_6$, which are normalized by the AdS curvature radius  $\ell$. 
The right figure shows an appropriate combination of the 5D entropy ${\mathcal S}_5$ and the temperature ${\mathcal T}$ as a function of the temperature. 
The thermodynamic relations for the 6D Schwarzschild black hole (dot-dashed line) and the 6D black string (solid line) are also plotted for a reference.
To generate the figures, three resolutions are used, $32 \times 128$ (cross), $64 \times 256$ (circle), and $128 \times 512$ (plus). 
}
\end{figure}

We have demonstrated the numerical method and shown numerical examples of the localized black holes. The largest localized black hole that we found is $\hatrho= \rhomax$. 
In this section, we try to understand the properties of the localized black holes. 
Let us first consider thermodynamic quantities of the localized black holes.  
Thermodynamic quantities of interest are the horizon temperature ${\cal T}$ and the entropy ${\cal S}_6$ of the black holes, which are given by 
\begin{eqnarray}
{\cal  S}_6 &=& \frac{ 1 }{4G_6}
  \left[ 
    2 \rho_0^4 \int d\Omega_3 \int^{ \pi/2}_0 
    d\chi ~\left( \frac{\ell}{z}\right)^4 e^{4B_0 + C_0}  { \sin^3 \chi } 
    \right]  ,
\\
{\cal T} &=&  
  \frac{1}{2\pi}  \frac{3}{2^{5/3} \rho_0}
          e^{A-B+C}|_{\rho=\rho_0}. 
\label{eq:T S}
\end{eqnarray}
In Fig. \ref{fig:S5T and S6T}, we plot a thermodynamic relation between these quantities  normalized by the AdS radius $\ell$.
The factor $G_6/\ell^4$ is introduced to obtain correctly rescaled quantity (see footnote \ref{footnote:1} in Sec. \ref{subsec:Conformal gauge method}).
Three resolutions are used to generate the figures. One sees that they are consistent and compatible with second-order scaling. In the figure, we have also plotted the same thermodynamic relations for the 6D Schwarzschild black hole and the 6D black string, for comparison  (see Appendix \ref{app:balck string}). 
One finds that for sufficiently small horizon radius $(T \gg \ell^{-1})$, the thermodynamic relation of the localized black hole is a 6D Schwarzschild black hole type. 
This behavior is confirmed for the 5D localized black holes \cite{Kudoh:2003vg}.
On the other hand, as the horizon radius increases $(T \ll \ell^{-1})$, the thermodynamic relation becomes a 6D black string type. 
The transition point at which the localized black hole changes its thermodynamic character is given by a temperature 
\begin{eqnarray}
    {\mathcal T} \approx \frac{1}{2\pi \ell}.
    \label{eq: transition T}
\end{eqnarray}
As could be expected, this temperature is given by the temperature of the localized black hole with $\hatrho = 1 $, which is ${\mathcal{T}}|_{\hatrho=1} \approx 0.9/2\pi\ell$. 
It is interesting that the horizon temperature $(\ref{eq: transition T})$ is the same as that of a ``critical'' black string whose horizon radius $r_h$ is $r_h=\ell$. 
Furthermore, the temperature reminds us of the Hawking-Page phase transition of the Schwarzschild-anti-de Sitter black hole \cite{Hawking:1983dh}.

We consider the specific heat,  ${\mathcal{T}} d{\mathcal{S}_6}/d{\mathcal{T}}$, of the localized black hole.  It is immediately clear from Fig. \ref{fig:S5T and S6T} that the specific heat is negative. 
Because the temperature ${\mathcal{T}}$ and the entropy ${\mathcal{S}_6}$ are well approximated by the 6D Schwarzschild black hole and the 6D black string in the respective regions $({\mathcal T} < \ell$ and ${\mathcal T} > \ell )$, the specific heat of the localized black hole is also well fitted by them.

There is an interesting quantity that characterizes the localized black hole by a geometrical quantity on the brane \cite{Kudoh:2003vg}.  
It is black hole ``entropy" measured by an observer living on the brane. We define the five-dimensional entropy by using proper circumference at $z=\ell$ as 
\begin{eqnarray}
 {\mathcal S}_5 = \frac{1}{4G_5}
 \left[ 2\pi^2 \rho_0 ^3  e^{3B_0 + 2C_0} 
 \right]_{z=\ell}
 \,.
\end{eqnarray}
In Fig. \ref{fig:S5T and S6T}, we display a relation between ${\mathcal S}_5$ and $\mathcal{T}$ taking an appropriate non-dimensional combination. In the figure, the same thermodynamic relations for the 6D black string and the 6D Schwarzschild black hole are also plotted for comparison.  
The figure shows that the five-dimensional entropy  $\mathcal S_5$ deviates from that of the 6D Schwarzschild black holes as the temperature decreases. 
Then the entropy $\mathcal S_5$ tends to behave like that of the 6D black string.  Because of limited resolution and computing power, we could not confirm whether $\mathcal S_5$ converges to the value of the 6D black string in the limit of large horizon radius (${\mathcal T} \to 0$).
However, an extrapolation of our results for $\mathcal S_5$ and $\mathcal S_6$ suggests that the thermodynamic relations of the localized black hole converge to those of the 6D black string in the limit. 
Recall that the gravity induced on the brane by the 6D black string is the 5D Schwarzschild black hole. 
Therefore, if the convergence is true, it means that the gravity induced on the brane by the large localized black holes is well described by the 5D Einstein gravity. 
Thus, the above results imply that the recovery mechanism works even in the presence of the black hole, which is not confined on the brane but localizes there, extending into the bulk.

We give a comment on a possible deviation from the thermodynamic character of the black string. 
For the weak gravity regime in the original 5D braneworld model, the relative order of the Kaluza-Klein corrections to Newton's law is $O(\ell^2/r^2)$  \cite{Garriga:1999yh,Sasaki:1999mi}, where $r$ is a coordinate radius. Similar Kaluza-Klein corrections will appear in the gravity induced on the brane by a black hole. 
However, the correction will be very small for massive black holes, and then numerical calculations can probe only the leading behavior. 
Therefore, even if numerical calculations show a good agreement between the thermodynamic character of the localized black hole and that of the black string, it does not mean that no correction to the 5D Schwarzschild black hole appears on the brane.

\begin{figure}
\centerline{
 \includegraphics[width=8cm,clip]{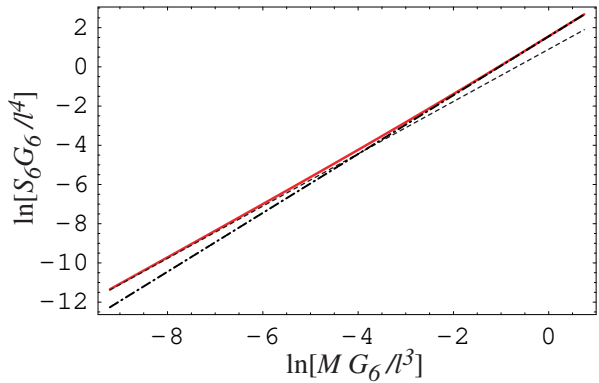}
}
\caption{ 
\label{fig:MassEntropy}
The relation between the thermodynamic mass and the entropy for the localized black holes (solid curve). 
The dashed and the dot-dashed lines represent the relations for the 6D Schwarzschild black hole and the 6D black string, respectively.
}
\end{figure}

Mass of a black hole is an important quantity to characterize the black hole. If one assumes the first law of thermodynamics \cite{Emparan:1999wa,Emparan:1999fd}, one can calculate a thermodynamic mass of a localized black hole. 
Because the localized black hole has the thermodynamic character analogous to those of the 6D Schwarzschild black hole and the 6D black string, the thermodynamic mass obtained by integrating the first law also agrees with them in the respective regime (Fig \ref{fig:MassEntropy}).
The mass of large black holes is then approximated by that of the 6D black strings, neglecting small corrections coming from the integration over small radii. It is given by 
\begin{eqnarray} 
 \mathcal{M} \approx \frac{1}{4} \left( \frac{3}{\sqrt{\pi}} S_6 \right)^{2/3} \left( \frac{\ell}{G_6} \right)^{1/3}  . 
\quad ( \hatrho \gtrsim 1)
\end{eqnarray}  
The black hole entropy $\mathcal S_6$ is greater than or nearly equal to that of the black string with the same mass.
This is contrary to the argument for the (BTZ-like) localized black holes and the BTZ black strings in Ref. \cite{Emparan:1999fd}.

\begin{figure}
\centerline{
 \includegraphics[width=8cm,clip]{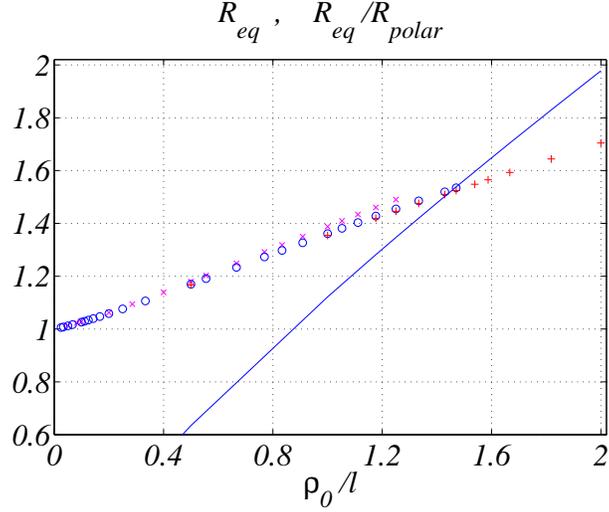}
}
\caption{ 
\label{fig:aspectratio}
Ratio of the equatorial to polar circumferential radii for the localized black holes $(\hatrho \lesssim  \rhomax)$. 
Three resolutions are used [$32 \times 128$ (cross), $64 \times 256$ (circle), and $128 \times 512$ (plus)].
The solid curve represents the equatorial radius $R_{eq}$ with respect to  $\hatrho$. 
}
\end{figure}

\subsection{Horizon geometry and curvatures}

\begin{figure}
\centerline{
 \includegraphics[width=8cm,clip]{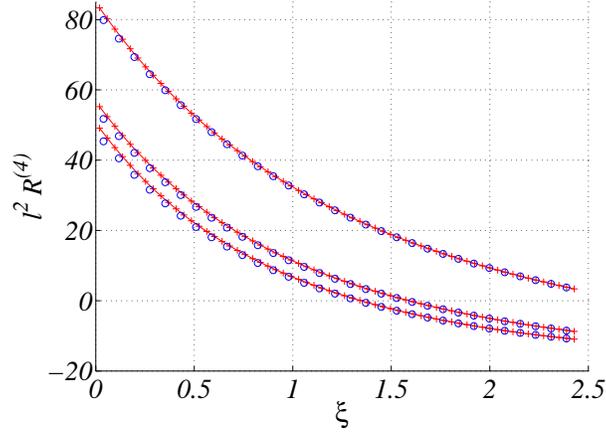}
}
\caption{ 
\label{fig:curvatureHorizon}
Intrinsic 4D curvatures $R^{(4)}$ of the horizon four-geometry for $\hatrho=0.5,~1, ~1.4$.  
The curve showing the largest positive curvature on the axis $\xi=0$ corresponds to $\hatrho=0.5$. As the radius increases, the curvature on the axis decreases. 
While the curvature remains positive for $\hatrho \lesssim 0.5$, the curvature becomes negative near the brane for $\hatrho \gtrsim 0.5$. 
}
\end{figure}

\begin{figure}
\centerline{
  \includegraphics[width=4.5cm,clip]{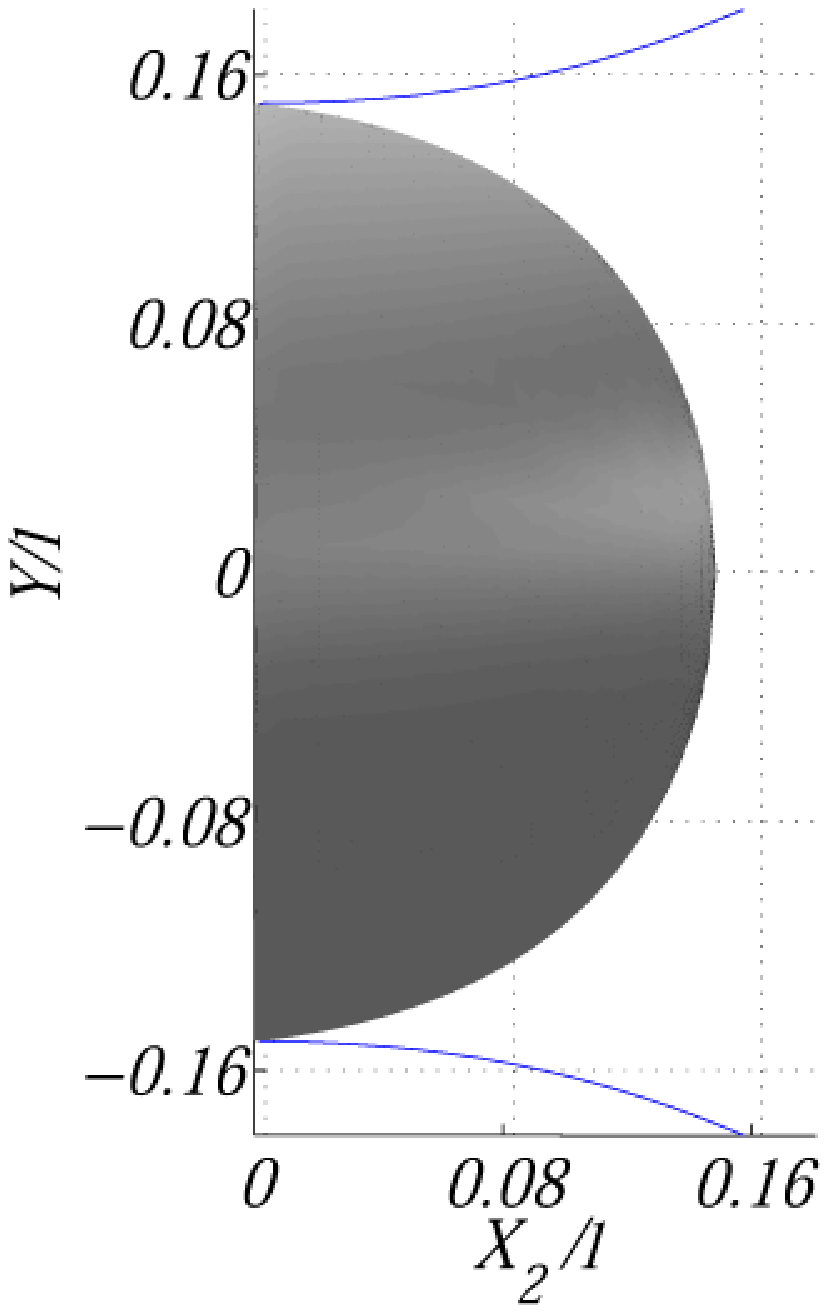}
\qquad
  \includegraphics[width=5.0cm,clip]{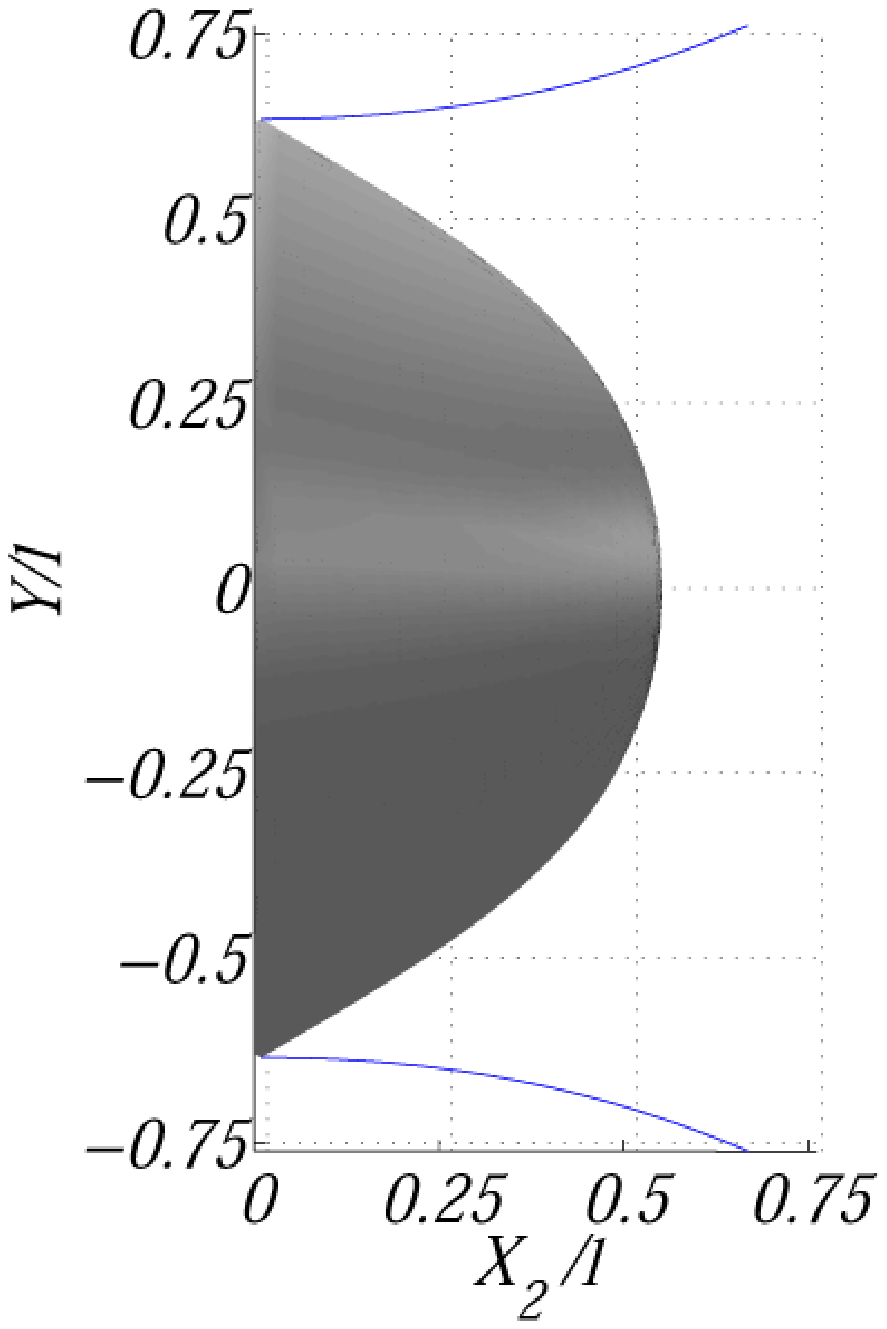}
}
\caption{
\label{fig:embedding geometry}
The four-geometry of the event horizon embedded in the Euclidean space (\ref{eq:euclidean space}), based on our numerical solutions. 
Two rotational degrees of freedom are suppressed. 
The size of the localized black holes in the left and right figures are $\hatrho = 0.1$ and $0.5$, respectively. 
The corresponding equatorial circumferential radii are $R_{eq}/\ell=0.15$ and $0.63$, respectively.
The solid curves show the location of the orbifold brane where the $Z_2$ symmetry is imposed in the original space.
} 
\end{figure}

A localized black hole is expected to have a flattened horizon geometry, and it will tend to flatten much more as its horizon radius increases ~\cite{Emparan:1999wa,Kudoh:2003vg,Kudoh:2003xz}. 
Let us define polar and equatorial circumferential radii of the horizon by 
\begin{eqnarray}
R_{eq} &=&  r  e^{B_0 + \frac{2}{3} C_0}|_{z=\ell} \,,
\nonumber 
\\ 
R_{polar} &=& \frac{4\rho_0}{2\pi} \int^{\pi/2}_{0} d\chi \left(\frac{\ell}{z} e^{B_0-C_0} \right) \,.
\end{eqnarray}
Figure \ref{fig:aspectratio} shows $R_{eq}/R_{polar}$ and $R_{eq}$. 
The small black holes are almost spherical, $R_{eq} \approx R_{polar}$, but the large black holes show flattened geometry, $R_{eq} > R_{polar}$.

The intrinsic curvature of the horizon geometry gives us another information. 
The horizon four-geometry is 
\begin{eqnarray}
 d\Sigma ^2 = 
 \frac{\ell^2}{z^2} \left[  e^{2(B_0-C_0)} \rho^2 d\chi^2 
 + r^2 e^{2B_0+4C_0/3} d\Omega^2_3   \right] \,.
\label{eq:horizon metric1}
\end{eqnarray}
The 4D intrinsic curvature $R^{(4)}$ of the four-geometry is shown in Fig. \ref{fig:curvatureHorizon} for several black holes.  
The curvature becomes negative near the brane for the black holes with $\hatrho \gtrsim 0.5$.   
The appearance of the negative spatial curvature near the brane is also confirmed for the EHM solution and the 6D black string (Appendix \ref{app:balck string}).  
Because of this negative spatial curvature, embeddings of the spatial horizon geometry into a five-dimensional Euclidean space 
\begin{eqnarray}
dE ^2 =  d{X}^2 + {dY}^2 + Y^2 d\Omega^2_3 
\label{eq:euclidean space}
\end{eqnarray}
are not possible for all black holes, but only possible for small black holes (see, e.g., Ref. \cite{MTW}).
The cylindrical and radial coordinates of the horizon are given by
\begin{eqnarray}
 X_{bh} (\chi) &=& 
  \int^{\pi/2}_{\chi} d\chi~
      \sqrt{ 
      \frac{ \rho_0^2\ell^2}{z^2}e^{2(B_0-C_0)}
- \left( \frac{dY_{bh}}{d\chi}\right)^2~     }  ,
\nonumber
\\
Y_{bh} (\chi) &=& \frac{\ell r}{z} e^{B_0 + \frac{2}{3} C_0} \,. 
\end{eqnarray}
Figure \ref{fig:embedding geometry} shows examples of the horizon geometry embedded in the Euclidean space. 
Because the cylindrical coordinate of the horizon on the axis does not coincide with the polar circumferential radius $R_{polar}$, the embeddings in the Euclidean space do not conserve the ratio $R_{eq}/R_{polar}$.  
To illustrate the horizon geometry, we rescale the cylindrical coordinate $X$ for each black hole to conserve the ratio. 
The rescaled cylindrical coordinate is denoted by $X_2$ (Fig. \ref{fig:embedding geometry}).
This illustration allows us to have an intuitive look of the horizon geometries for different horizon radii.

\begin{figure}
\centerline{
  \includegraphics[width=8.0cm,clip]{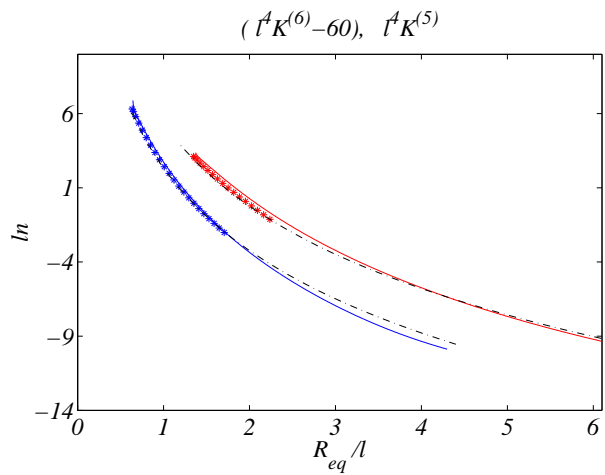} 
}
\caption{  
\label{fig:curvature}
 6D and 5D squares of Riemann curvature, $\ln (\ell^4 K^{(6)}-60)$ (asterisk) and $\ln( \ell^4 K^{(5)})$ (solid curve), on the brane for $\hatrho = 0.5,~1.25$. 
The dot-dashed curves show fits to the asymptotic behavior (\ref{eq:K5 propto R8}).  
}
\end{figure}

While we have compared the localized black hole to the black string, the black string has a curvature singularity at the AdS horizon as well as the ordinary singularity at the center of the event horizon (Appendix \ref{app:balck string}) \cite{Chamblin:1999by}.  Thus the solution has been considered as an unphysical solution.  Do the localized black holes have a naked singularity somewhere?  To test the regularity of the spacetime, we calculate the square of Riemann tensor that is one of curvature scalars, 
\begin{eqnarray}
    K ^{(6)} := R^{\mu\nu\lambda\rho} R_{\mu\nu\lambda\rho}. 
\end{eqnarray}
It is hard to numerically calculate this kind of curvature scalars with enough accuracy near the axis and the horizon. In particular near the axis ($r=0$), we encounter severe cancellations between terms that have coefficients proportional to $1/r^4$, $1/r^3$, $\cdots$.  Nevertheless, we can confirm that the curvature  $K^{(6)}$ is finite with good behavior. Then as far as the curvature $K^{(6)}$ is concerned, localized black holes have no naked singularity, at least in the computational domain. 
This supports that our localized black holes are physically acceptable black hole solutions.   
In Fig. \ref{fig:curvature}, we illustrate the curvature $K^{(6)}$ on the brane. It goes rapidly to a constant value given by the AdS$_6$, 
\begin{eqnarray}
    K^{(6)}  = \frac{60}{\ell^4} , \quad ({AdS_6})  
\end{eqnarray}
and then it begins to fluctuate due to roundoff error. 
Thus in Fig. \ref{fig:curvature}, we show $K^{(6)}$ for an appropriate range of proper circumferential radius $R_{eq}$ on the brane.
We have furthermore calculated the square of the 5D Riemann tensor 
\begin{eqnarray}
K^{(5)} :=  R^{ {(5)}\mu\nu\lambda\rho} R^{(5)}_{\mu\nu\lambda\rho} , 
\label{eq:K5}
\end{eqnarray}
by using the induced metric on the brane.  This curvature is a good measure for an observer on the brane.   
The scalar curvature for the 5D Schwarzschild black hole is 
\begin{eqnarray}
    K^{(5)} \propto  \frac{1}{ R_{eq}^8 } \,,
\label{eq:K5 propto R8}
\end{eqnarray}
at a distance from the horizon [see Eq. (\ref{eq:string K5})]. 
On the other hand, the curvature $ K^{(5)} $ for the 6D Schwarzschild black hole is proportional to $ \sim { 1/R_{eq}^{10} }$.  We could then compare these behaviors with that of the curvature for the localized black holes. 
We plot $K^{(5)}$ of the localized black holes with $\hatrho=0.5$ and $1.25$ in Fig. \ref{fig:curvature}. 
The dot-dashed curves are fits for the form (\ref{eq:K5 propto R8}) in an appropriate asymptotic region. 
Although the localized black hole with $\hatrho=1.25$ is not so large, we observe that the 5D curvature for $\hatrho=1.25$ shows better agreement with Eq.  (\ref{eq:K5 propto R8}), compared to that for the smaller black hole. We note that the fit to the behavior (\ref{eq:K5 propto R8}) is better than a fit to the behavior $\propto { 1/R_{eq}^{ 10} }$. 
However we refrain from going into a discussion of such asymptotic behavior, because the asymptotic boundary conditions adopted in this paper are too naive to discuss it.

\section{Conclusion}
\label{eq:conclusion}

No realistic black holes localized on the brane have been found so far in the warped compactification, except for the EHM solutions in the four-dimensional braneworlds. 
Black holes are important physical objects to understand and to test strong gravitational phenomena in the warped compactification. 
We have explored this problem by means of numerics, constructing black holes localized on the brane. 
We have employed the conformal gauge method that has been developed to study relativistic stars on the brane, nonuniform black strings and Kaluza-Klein black holes. 
In our previous work, we could not succeed in constructing large black holes, but could construct only the small black holes ($\hatrho  \lesssim 1/5 $). 
However, improved numerics allow us to construct the large localized black holes with $\hatrho  \gtrsim 1$, and then we have studied the properties of the localized black holes.

The largest size of the black holes that could be relaxed in a reasonable computational time is $\hatrho = \rhomax$ in this paper. The parameter $\hatrho$ is approximately equal to the proper circumferential radius on the brane. 
The largest size depends on the resolution of the calculation, and thus it is limited by the computational time. 
The situation would be similar to the study of the large relativistic stars \cite{Wiseman:2001xt}.

The horizon geometry of the localized black holes is flattened in the bulk. 
The ratio of the equatorial to polar circumferential radii shows that the localized black hole becomes much more flattened as its horizon radius increases (Fig. \ref{fig:aspectratio}). 
We have pointed out that the intrinsic curvature of the horizon four-geometry becomes negative near the brane for the black holes with $\hatrho \gtrsim 0.5$. Then the global shape of the flattened horizon changes as its horizon radius increases.

The thermodynamic properties of the localized black holes have been discussed by investigating the entropy, temperature, and specific heat of the black holes. 
For the small black holes ($\hatrho \lesssim 1 $) the thermodynamic relation between the entropy and the temperature is the 6D Schwarzschild black hole type, whereas for the large black holes ($\hatrho  \gtrsim 1$) the thermodynamic relation is the 6D black string type (Fig. \ref{fig:S5T and S6T}). 
The transition occurs at ${\mathcal T} \approx 1/2\pi \ell$, corresponding to $\hatrho \approx 1$.
The 5D entropy measured by an observer on the brane also shows the transition: the physical quantities of the large localized black holes are approximately described by the 6D black strings. 
Thus the large localized black hole looks like the 6D black string with its ends capped off.
This transition of thermodynamic relations might be analogous to an expected transition between a sequence of (static) black string solutions and that of black hole solutions in the Kaluza-Klein compactification~\cite{Kol:2002xz}, although the transition is questioned by a recent study (Ref. \cite{Kudoh:2003ki}).

Recall that the induced gravity of the 6D black string is given by the 5D Schwarzschild black hole. Then the above results indicate that the gravity induced on the brane by the large localized black hole is approximated by the 5D Einstein gravity. 
Such a recovery mechanism of lower-dimensional Einstein gravity on the brane is proved in the weak gravity regime by means of linear and second-order perturbations \cite{Garriga:1999yh,Kudoh:2001wb}, and it is partially confirmed for the strong gravity regime by the numerical study \cite{Wiseman:2001xt}. 
Our result also suggests that the recovery mechanism works even in the presence of the black hole that is bound to the brane.

Let us discuss the relation between our results and the argument of the classical black hole evaporation conjecture \cite{Tanaka:2002rb,Emparan:2002px}. 
The argument claims that large localized \textit{static} black holes may not exist. 
On one hand, the physical quantities calculated in this paper have not indicated a critical point at which a sequence of solutions suddenly disappears or some kind of singular behavior appears. 
Thus it seems to shed doubts on the conjecture. 
On the other hand, however, the largest localized black hole in this paper is $\hatrho = \rhomax$, and the black hole is not sufficiently large. It would be still small compared with the black holes supposed in the conjecture. 
It is certainly true that we cannot provide a conclusive disproof (or proof) of the conjecture by means of numerics, even if we construct a very large black hole with, e.g., $\hatrho =100$. 
In Ref. \cite{Kudoh:2003vg}, some possible consistent scenarios are discussed. 
However, we expect that the sequence of our localized black holes represents the unique stable vacuum solution without the naked singularity that describes a nonrotating uncharged black hole bound to the brane.
Needless to say, it is preferable to have an analytic solution of a localized black hole, and thus much progress in the study of braneworld black holes is required.

An interesting application of the conformal gauge method is a study of black holes in the RS two-branes model. Because the localized black holes in the model are similar to black holes in the Kaluza-Klein compactification, the method in Ref. \cite{Kudoh:2003ki} can be directly applied. 
Effects of the warped compactification, the second brane, and radius stabilization mechanism on the black holes would result in different phenomena observed in this paper. 
Charged localized black holes are also interesting. Introducing gauge fields on the brane is not difficult in our approach. 
We hope to report soon our results toward these directions.

\begin{acknowledgments}
The author would like to thank Hideo Kodama, Takahiro Tanaka, Toby Wiseman, Shinji Mukohyama, and Bruce Bassett for their valuable comments and discussions. 
The author is grateful to Toby Wiseman for his hospitality at Cambridge.
The numerical computations reported in this work were carried out at the Yukawa Institute Computer Facility and at the NAO.
This work is supported in part by the JSPS and a Grant-in-Aid for the 21st Century COE.
\end{acknowledgments}

\appendix 

\section{6D black string}
\label{app:balck string}
The black string solution in the single brane model has a singularity at the AdS horizon $(z\to \infty)$ \cite{Chamblin:1999by}. However, the solution is still useful as a reference.  The 6D black string is given by 
\begin{eqnarray}
ds^2 =\frac{\ell^2}{z^2} 
\left[    - V dt^2 + V^{-1} dr^2 +r^2  d\Omega_3^2 +dz^2 
\right]\,,
\end{eqnarray} 
where $V= 1- {r_h^2}/{r^2} $. The brane is at $z=\ell$.
The square of the Riemann tensor is 
\begin{eqnarray}
K^{(6)} = R_{\mu\nu\lambda\rho} R^{\mu\nu\lambda\rho}
=
\frac{1}{\ell^4} \left[ 60 +  \frac{72 r_h^4z^4}{r^8} \right], 
\end{eqnarray}
which diverges at the AdS horizon ($z\to \infty$), as well as at the black string singularity at $r=0$.  The 5D curvature scalar (\ref{eq:K5}) on the brane is given by 
\begin{eqnarray}
K^{(5)} = 72 r_h^4/r^8.
\label{eq:string K5}
\end{eqnarray}

The intrinsic curvature $R^{(4)}$ of the horizon four-geometry is 
\begin{eqnarray}
	R^{(4)} = \frac{6}{r_h^2 \ell^2} (z^2-2r_h^2).
\end{eqnarray}
This is not always positive, but becomes negative for $ z < \sqrt{2} r_h$.

The temperature and entropy of the black string are 
\begin{eqnarray}
 {\mathcal T} &=& \frac{1}{2\pi r_h} , 
\cr
{\mathcal S_6} &=&  \frac{1}{4G_6}\left[ \frac{(2\pi)^2}{3} r_h^3 \ell \right].
\end{eqnarray}
Assuming the first law, thermodynamic mass is given by 
\begin{eqnarray*}
 \mathcal{M} =  \frac{1}{4} \left( \frac{3}{\sqrt{\pi}} S_6 \right)^{2/3} \left( \frac{\ell}{G_6} \right)^{1/3}  .
\end{eqnarray*}

\section{Asymptotic boundary conditions}
\label{sec:asymptotic bc}

\subsection{Linear perturbations}

The asymptotic behavior of black hole characterizes the black hole itself, and then it is important. To impose physically reasonable asymptotic boundary conditions, we need to solve linear perturbations of the black hole geometry in the asymptotic region.
Here we assume that the asymptotic behavior of black hole geometry is equivalent to that of gravitational field induced by a point mass on the brane. 
Then we consider solutions to the linear perturbations of weak gravitational field in the conformal gauge. 
Linearized equations in the conformal gauge do not yield decoupled equations, and then it is difficult to solve it.  
However, using the transverse-traceless gauge, one finds that a master equation exists and it is solved using a Green's function.  
The discussion in this appendix is based on an argument in Ref.~\cite{Wiseman:2001xt}.

Perturbations in the transverse-traceless (TT) gauge are given by 
\begin{eqnarray}
ds^2 
 &=& \frac{\ell^2}{z^2} \left[ 
         -e^{2F}dt^2 + e^{2H}dr^2 + r^2 e^{2D} d\Omega +  dz^2 
     \right], 
\end{eqnarray}
where $F$ and $D$ are determined by the TT conditions,
\begin{eqnarray}
    F = -\frac{1}{r^3} \partial_r (r^4 H) \,, 
\quad
    D = \frac{1}{3r^2} \partial_r (r^3 H) .
\end{eqnarray}
The linearized equation for $H$ is given by 
\begin{eqnarray}
0  = \frac{z^2}{\ell^2} \left[ \partial_r^2 +\frac{5}{r}\partial_r
       +  \partial_z^2 - \frac{4}{z}\partial_z 
   \right] H \,.
\label{eq:H eq}
\end{eqnarray}
Comparing these coordinates to those in the conformal gauge, coordinate transformations to bring the metric into the conformal gauge are $r'_{TT} = r + f $, $z'_{TT} = z + g $
with 
$g_{,z} = f_{,r} +H \,, g_{,r} = -f_{,z} $
which yield Poisson equations for $f$ and $g$ \cite{Wiseman:2001xt}, 
\begin{eqnarray}
 \triangle  g  = H_{,z}  \,,
\quad 
 \triangle f   = - H_{,r} \,.
\label{eq:poisson f, g}
\end{eqnarray}
The linearized equation (\ref{eq:H eq}) does not change by the coordinate transformations.  
Then the coordinate transformed metric components are given in terms of $H$, $f$, and $g$:
\begin{eqnarray}
A_0  &=& - \frac{g}{z} + F  \,,
 \nonumber
\\
B_0 &=&    \frac{3f}{5r} - \frac{g}{z} + \frac{2}{5}\partial_z  g 
    + \frac{3}{5} D \,,
\nonumber
\\
C_0  &=&  \frac{3f}{5r} - \frac{3}{5}g_{,z} +\frac{3}{5} D \,.
\label{eq:asymptotic ABC}     
\end{eqnarray}

To solve the coordinate transformations (\ref{eq:poisson f, g}) for $f$ and $g$, we need to specify the boundary conditions. 
Boundary conditions on the brane are derived from the brane matching condition that is expanded to linear order,
\begin{eqnarray}
  \partial_\rho^2 g  +  \frac{3}{\rho} \partial_\rho g  &=& 0 \,,
\cr
  3 \partial_\rho g &=& H_{,\chi}  \,,
\end{eqnarray}
and we also have $3 \partial_\chi f =  \partial_\chi  H $. On the axis, we impose 
\begin{eqnarray}
f=0.
\end{eqnarray}

The asymptotic behavior (\ref{eq:asymptotic ABC}) of gravitational field induced by a point particle can be used as asymptotic boundary conditions of black hole geometry. The boundary conditions are given by Neumann conditions, eliminating an overall coefficient that corresponds to a mass of a black hole. To give the boundary conditions (\ref{eq:asymptotic ABC}), we must first solve Eq. (\ref{eq:H eq}) and then determine the coordinate transformations by solving the Poisson equations (\ref{eq:poisson f, g}). 
 In the next subsection, we discuss a solution to the linearized equation (\ref{eq:H eq}) and the coordinate transformations.

\subsection{Solution  to linear perturbations }

Linearized gravity in the RS background has been discussed several times in the literature. 
A treatment of it in general dimensions is discussed in Ref. \cite{Giddings:2000mu}. In this reference, metric perturbations in the transverse-traceless gauge are solved by means of a Green's function.
While the solution is general, analytic estimations of $H$ and the coordinate transformations become somewhat messy. Thus to evaluate the leading order contributions, we consider another method that is rather simple.

A solution of (\ref{eq:H eq}) is constructed as 
\begin{eqnarray}
H= \ell \sqrt{\frac{2}{\pi}}   
\left(\frac{z}{\ell} \right)^{5/2}  
\Bigl( \frac{ \ell }{r} \Bigr)^2   
\int^{\infty}_{0}  dk \left[
 (k \ell)^{3/2}
h(k) 
 J_{2}(kr) 
 K_{5/2}(kz)
\right] \,,
\end{eqnarray}
where $J_2$ and $K_{5/2}$ are the Bessel function and modified Bessel function, respectively, and $h(k)$ is an arbitrary function.
Because we are interesting in asymptotic behavior ($r \gg \ell$ or $z\gg \ell$), only long-wave modes give dominant contribution in the integration. Let us consider $h(k)= {\mathrm{const.}}=h_0$. Then we obtain an approximate solution,  
\begin{eqnarray}
H= 
\frac{h_0 \ell^2}{r^2} 
\left( 
     \frac{3}{2}- \frac{z^2}{r^2} 
   + \frac{z^5}{r^2 (r^2+z^2)^{3/2}}
\right).
\label{eq:sol H}
\end{eqnarray}

Particular solutions of Eq. (\ref{eq:poisson f, g}) for the asymptotic behavior (\ref{eq:sol H}) can be found easily by using polar coordinates. They are 
\begin{eqnarray}
g_{p} &=&  \frac{ h_0 \ell^2 
    \left( -8 r^4 - 4 r^2 z^2 + z^4 - 
      z {\sqrt{r^2 + z^2}} \left( 3 r^2 + z^2 \right) 
      \right) }{3 r^2 
    {\left( r^2 + z^2 \right) }^{\frac{3}{2}}} \,,
\cr
f_{p} &=&  \frac{ h_0 \ell^2}{6 r^3}  \left( 11 r^2 - 2 z^2 + \frac{4 r^4}{r^2 + z^2} + 
      \frac{2 z \left( -8 r^4 - 4 r^2 z^2 + z^4
           \right) }{{\left( r^2 + z^2 \right) }^
         {\frac{3}{2}}} \right) \,.
\end{eqnarray}
To obtain solutions that satisfy required boundary conditions, we need to add homogeneous solutions. Then general solutions are given by 
\begin{eqnarray}
g(r,z) = g_p + \int^{\infty}_{0} d q F(q) \cos (q r/\ell) e^{- q z/\ell},
\cr
f(r,z) = f_p - \int^{\infty}_{0} d q F(q) \sin (q r/\ell) e^{- q z/\ell}.
\label{eq:sol f,g}
\end{eqnarray}
From the boundary condition $3 \partial_r g = - r \partial_z H$ on the brane, we obtain 
\begin{eqnarray}
F(q) & =& \frac{2 \ell  h_0  }{3} +  \frac{8h_0 \ell  e^{q}}{3\pi} 
    \left[ 2K_{0}(q) - qK_{1}(q) \right].
\end{eqnarray}

On the brane $(z=\ell, r \gg \ell) $, the leading order contributions are evaluated as 
\begin{eqnarray}
g  &\approx&    - \frac{\ell^3 h_0}{3r^2},  \quad  
\cr
f  &\approx & \frac{\ell^2 h_0}{ r} O(\log(r/\ell) ). 
\end{eqnarray}
Near the axis ($z\gg \ell$, $r\ll \ell $), we have
\begin{eqnarray}  
g  &\approx&  O\left(\frac{\log z}{z} \right), 
\nonumber
\\
f  &\approx& \frac{r h_0}{z} \Bigl[ O\left(\frac{1}{z}\right)+ O\left(\frac{\log z}{z}\right)\Bigr]  .
\end{eqnarray}   
The coordinate transformation (\ref{eq:sol f,g}) and the asymptotic behavior (\ref{eq:sol H}) (or the solution in Ref. \cite{Giddings:2000mu}) with Eq. (\ref{eq:asymptotic ABC}) can be used as the asymptotic boundary condition of localized black holes.


\bibliographystyle{apsrev} 

\end{document}